\documentclass[11pt,a4paper]{article}
\usepackage{latexsym}
\usepackage{graphicx}
\usepackage[all]{xypic}
\usepackage{epsf}
\usepackage{color}
\usepackage{enumerate}

\usepackage{algorithm}
\usepackage{algpseudocode}

\usepackage{amssymb}

\newtheorem{theorem}{Theorem}
\newtheorem{proposition}[theorem]{Proposition}
\newtheorem{lemma}[theorem]{Lemma}
\newtheorem{corollary}[theorem]{Corollary}
\newtheorem{definition}{Definition}

\newtheorem{remark}{Remark}
\newtheorem{fact}{Fact}
\newtheorem{example}{Example}

\newenvironment{proof}{\noindent\textit{Proof. \  }}{\hfill$\Box$\medskip\par}

\newcommand{\N}{\mathbb{N}}
\newcommand{\R}{\mathbb{R}}

\renewcommand{\emph}[1]{\textsl{#1}}

\newcommand{\calF}{{\cal F}}

\newcommand{\calH}{{\cal H}}

\newcommand{\calS}{{\cal S}}

\newcommand{\st}[2]{\ensuremath{\left\langle \mathit{#1} ; #2 \right\rangle}}
\newcommand{\su}[2]{\ensuremath{#1_{1}, \ldots, #1_{#2}}}

\newcommand{\raus}[1]{}
\newcommand{\pb}[1]{\textsc{#1}}

\newcommand{\np}{\mathrm{NP}}

\def\ACQ#1{\textbf{ACQ}{#1}}
\def\ACQineq#1{\textbf{ACQ}^{\neq}{#1}}
\def\ACQplus#1{\textbf{ACQ}^{+}{#1}}

\def\strictACQ#1{\textbf{ACQ}_1{#1}}
\def\strictACQineq#1{\textbf{ACQ}^{\neq}_1{#1}}
\def\strictACQplus#1{\textbf{ACQ}^{+}_1{#1}}

\def\FACQ#1{\textbf{F-ACQ}{#1}}
\def\FACQineq#1{\textbf{F-ACQ}^{\neq}{#1}}

\def\FACQplus#1{\textbf{F-ACQ}^{+}{#1}}

\def\strictFACQ#1{\textbf{F-ACQ}_1{#1}}
\def\strictFACQineq#1{\textbf{F-ACQ}^{\neq}_1{#1}}

\def\strictFACQplus#1{\textbf{F-ACQ}^{+}_1{#1}}

\def\FFO#1{\textbf{F-UFO}{#1}}
\def\FFO#1{\textbf{F-FO}{#1}}

\def\FAFO#1{\textbf{F-AFO}{#1}}

\def\strictFAFO#1{\textbf{F-AFO}_1{#1}}

\def\FFOvar2#1{\textbf{F-FO}^{var 2}{#1}}
\def\strictFFOvar2#1{\textbf{F-FO}^{var 2}_1{#1}}

\def\tu#1{\mathbf{#1}}

\newcommand {\ssi}{\leftrightarrow}
\newcommand {\imp}{\rightarrow}
\newcommand {\et}{\wedge}\newcommand {\Et}{\bigwedge}
\newcommand {\ou}{\vee}\newcommand {\Ou}{\bigvee}
\newcommand {\fa}{\forall}

\newcommand {\var}[1]{\mbox{var}(#1)}

\newcommand {\arity}[1]{\mbox{arity}(#1)}

\newcommand {\sg}{\sigma}
\renewcommand{\t}{\tau}

\begin{document}

\fontencoding{OT1}
 
\fontfamily{ppl}
 
 
 

\title{The complexity of acyclic conjunctive queries revisited}

\author{
Arnaud Durand~\thanks{\'Equipe de Logique Math\'ematique - CNRS UMR 7056.
  Universit{\'e} Denis Diderot - Paris~7,
   2 place jussieu, 75251 Paris cedex 05, France.
Email : \texttt{durand@logique.jussieu.fr}}
  \and 
  Etienne Grandjean~\thanks{GREYC - CNRS UMR 6072.
 Universit\'e de Caen, 14032 Caen, France
 Email : {\tt grandjean@info.unicaen.fr}}
}

\date{}

\maketitle


\bigskip
\begin{center}
\end{center}

\begin{abstract}
In this paper, we consider  first-order logic over unary functions and study the  complexity of the evaluation problem for conjunctive queries described by such kind of formulas.

A natural notion of query acyclicity for this
language is introduced and we study the complexity of a large number of variants or
generalizations of acyclic query problems in that context (Boolean
or not Boolean, with or without inequalities, comparisons,
etc...). Our main results show that all those problems are \textit{fixed-parameter linear} i.e. they can be
evaluated in time $f(|Q|).|\textbf{db}|.|Q(\textbf{db})|$ where
$|Q|$ is the size of the query $Q$, $|\textbf{db}|$ the database size,
$|Q(\textbf{db})|$ is the size of the output and $f$ is some
function whose value depends on the specific variant of the query
problem (in some cases, $f$ is the identity function).

Our results have two kinds of consequences. First, 
they can be easily
translated  in the relational (i.e., classical) setting. Previously known bounds for some query problems are improved and new tractable cases are then exhibited. Among
others, as an immediate corollary, we improve a result of
~\cite{PapadimitriouY-99} by showing that any (relational) acyclic
conjunctive query with inequalities can be evaluated in time
 $f(|Q|).|\textbf{db}|.|Q(\textbf{db})|$.
 
 A second consequence of our method is that it provides a very natural descriptive approach to the complexity of well-known algorithmic problems. A number of examples (such as acyclic subgraph problems, multidimensional matching, etc...) are considered for which new insights of their complexity are given. 
\end{abstract}

\newpage

\tableofcontents

\newpage

\section{Introduction}

\raus{blabla - mesure de complexité - commentaire sur pb de
requêtes - cas tractables - presentation du papier}

The complexity of relational query problems is an important and
well-studied field of database theory. In particular, the class of
conjunctive queries (equivalent to select-project-join queries)
which are among the most simple, the most natural and the most
frequent type of queries have received much attention.

A query problem takes as input a database $\textbf{db}$ and a
query $Q$ and outputs  $Q(\textbf{db})$ the result of the
evaluation of $Q$ against $\textbf{db}$ (when the query is
Boolean, $Q(\textbf{db})$ is simply \textit{yes} or \textit{no}).
There exist mainly two ways to investigate the complexity of such
a problem. In the \textit{combined complexity} setting, one
expresses the complexity of the problem in terms both of the
database size $|\textbf{db}|$ and of the query size $|Q|$ (and of
the output size $ |Q( \textbf{db})| $ if necessary). It is
well-known that, in that context, the Boolean conjunctive query
problem is $\np$-complete (\cite{ChandraM-77, AbiteboulHV-95}).
However, it is natural to consider that the database size is
incomparably bigger than the query size and to express the
complexity of the problem in terms of the database size only. In
that case, the complexity of the conjunctive query problem falls
down to $P$ (and even less). However, as discussed by
~\cite{PapadimitriouY-99}, that point of view is not completely
satisfactory because although the problem becomes polynomial time
decidable, the formula size may inherently occur in the exponent
of the polynomial. Even for small values of this parameter, this
may lead to non tractable cases.

An interesting notion from parameterized complexity
(\cite{DowneyF-99}) that appears to be  very useful in the context
of query evaluation (see~\cite{PapadimitriouY-99}) is fixed
parameter tractability. A (query) problem is said to be
\textit{fixed-parameter (f.p.) tractable (resp. linear)} if its time complexity
is $f(|Q|).P(|\textbf{db}|,|Q(\textbf{db})|)$ for some function
$f$ and some polynomial $P$ (resp. linear polynomial $P$). In that case, the formula size
influences the complexity of the problem by a multiplicative
factor only. Identifying the fragments of relational queries that
are f.p. tractable for small polynomials $P$ is then an important
but difficult task. Surprisingly, a very broad and well-studied
set of queries appears to lie within this class: as shown in
\cite{Yannakakis-81} (see also~\cite{FlumFG-02} for a precise
bound), $\ACQ{}$, the \textit{acyclic} conjunctive query problem
(we refer to the standard notion of acyclicity in databases; for
precise definitions see section~\ref{SEC Definitions of query
problems}) can be solved in polynomial time
$O(|Q|.|\textbf{db}|.|Q(\textbf{db})|)$. Besides, it has been
proved that evaluating an acyclic conjunctive query is not only
polynomial for sequential time but also highly parallelizable
(see~\cite{GottlobLS-01}).

 A natural extension $\ACQineq{}$ of $\ACQ{}$ allows inequalities between variables,
  i.e., atoms of the form $x\neq y$.  In~\cite{PapadimitriouY-99}, it is shown that this latter
class of queries is also f.p. tractable and can be evaluated in
time $g(|Q|).|\textbf{db}|.|Q(\textbf{db})|. \log^2 |\textbf{db}|$
where $g$ is an exponential function. Despite of these results, a
lot of query problems including the  extension of acyclic
queries obtained by allowing comparisons of the form $x<y$ are
likely f.p. intractable as shown again by
~\cite{PapadimitriouY-99}.

In this paper,
we revisit the complexity of acyclic conjunctive
queries under a different angle. First, a class of
so-called \textit{unary functional queries} based on first-order
logic over unary functions is introduced. 
 Focusing on
the existential fragment of this language, we introduce a very
natural
 graph-based notion of query acyclicity.
 We then show that various classes of relational conjunctive query
problems can be easily interpreted in \textit{linear time}  by
corresponding  (unary) functional conjunctive query problems (see
section~\ref{SEC From CQ to FQ}): this is done by switching from
the classical language describing relations between elements of
some domain $D$ (i.e., the relational setting) to a functional one
over the universe of tuples: unary functions basically describe
attribute values. In this context, unary functional formulas can be seen as a logical
embodiment of the well-known \textit{tuple calculus}. A nice feature of the reduction is that it
preserves acyclicity of queries in the two different contexts. The
main part of the paper (section~\ref{SEC The complexity of
generalized acyclic queries}) is devoted to the analyze of the
complexity of the query problem for a wide range of syntactically
defined functional formulas. More precisely, whether inequalities ($\neq$)
are allowed or not, whether the query is Boolean or not or whether a  restricted use of comparisons ($<$) is allowed are considered.
In each case, we show that such queries can be evaluated in time
$f(|Q|).|\textbf{db}|.|Q(\textbf{db})|$ (in time
$f(|Q|).|\textbf{db}|$ for the Boolean case)  where the value of
function $f$ depends on the precise (functional) query problem
under consideration.

Coming back to the relational setting,  as immediate corollaries, we obtain a
substantial (and optimal) improvement of the bound proved
in~\cite{PapadimitriouY-99} for the $\ACQineq{}$ problem and a new
proof of the complexity of the $\ACQ{}$ problem. Moreover, we
generalize the complexity bound for $\ACQ{}$ to a slightly larger
class of queries denoted by $\ACQplus{}$ that allow comparisons
($<$, $\leq$, $\neq$) in a restricted way. This should be compared with the result of~\cite{PapadimitriouY-99} which shows that an unrestricted use of comparisons inside formulas leads to an intractable query problem. 
The results of this paper implies that, regardless of the query size,
 $\ACQ{}$, $\ACQplus{}$
and $\ACQineq{}$ are inherently of the same \textit{data}
complexity. 

\medskip

One can easily describe algorithmic problems by queries written in some language. This allows to reduce the complexity of these problems to the complexity of query evaluations for the language. 
In section~\ref{SEC Fixed-parameter linearity of some natural problems}, this well-known descriptive approach is used for a number of algorithmic problems (like acyclic subgraph isomorphism, multidimensional matching, etc..). They are considered as well in their decision version as in their function  or enumeration (of solutions) version.  The variety of languages considered in the paper 
permits to express easily (i.e. without encoding) a large kind of properties (on graphs, sets, functions, etc...). Our results provides new insight on the complexity of these problems. In all cases, the best known (data) complexity bounds  is at least reproved and sometimes improved.  

\medskip

The methods we use to prove the main results of this
paper are, as far as we know, original and quite different from
those used so far in this context. They are essentially a
refinement of the methods introduced in
 a recent technical report by Fr\'ed\'eric Olive and the present authors (see~\cite{DurandGO-04}): that paper essentially deals with hierarchies of definability inside existential second order logic in connection with nondeterministic linear time.
 As~\cite{DurandGO-04} did before, we introduce here a simple combinatorial notion on unary
functions called \textit{minimal sample} (see section~\ref{SEC
Samples of unary functions}) and develop over this notion some technics of quantifier elimination in formulas that can be performed in linear time.
Considering unary functions in the language permits the introduction of simple but powerful new logico-combinatorial methods (based on graphs mainly). Arguments for this are given here  through the consequences on the complexity of relational acyclic conjunctive queries. There are possible other applications of the methods and the language; they are discussed in the conclusion.


\section{Preliminaries}~\label{SEC Preliminaries}

The reader is expected to be familiar with first-order logic (see
e.g.~\cite{EbbinghausF-99, Libkin-04}) but we briefly give some
basic definitions on signatures, first-order structures and
formulas.

 A {\em signature} (or {\em vocabulary}) $\sigma$ is a finite set of
relation and function symbols, each of which has a fixed arity
which can be zero (0-ary function symbols are constant symbols and
0-ary relation symbols are Boolean variables). The {\em arity} of
$\sigma$ is the maximal arity of its symbols.
 A signature whose arity is at most one is said to be \textit{unary}.

A (finite) {\em structure} $\calS$ of vocabulary $\sigma$, or
$\sigma$-structure, consists of a finite domain $D$ of cardinality
$n>1$, and, for any symbol $s\in\sigma$, an interpretation of $s$
over $D$ (often denoted also by $s$, for simplicity). \raus{The union
of two structures $\calS$ and $\calS'$ over the same domain but on
disjoint signatures will be denoted $(\calS,\calS')$.}

We will often deal with {\em tuples} of objects. We denote them by
bold letters: for example, $\tu x = (x_1,\dots,x_k)$. If $\tu f$
is a $k$-tuple of functions $(f_1,\dots,f_k)$, then $\tu f(x)$
stands for $(f_1(x),\dots,f_k(x))$. Analogously, if $\tu f$ and
$\tu g$ are two $k$-tuples of functions, $\tu f(x) = \tu g(y)$
stands for the logical statement: $f_1(x) = g_1(y) \wedge \dots
\wedge f_k(x)=g_k(y)$.

  Let $\varphi \equiv \varphi(x_1,\dots, x_k)$ be a
first-order formula of signature $\sigma$ and free variables among
$x_1,\dots,x_k$. Let $\var{\varphi}$ denote
the set of variables of $\varphi$ .  Let $\mathcal{L}$ be a class of
first-order formulas (also called a query language). The
\textit{query} problem associated to $\mathcal{L}$ (and also
denoted by $\mathcal{L}$) is defined as follows:

\medskip

\noindent \textbf{Input:} A signature $\sg$, a $\sg$-structure
$\calS$ of domain $D$ and a
first-order $\sg$-formula $\varphi(x_1,\dots, x_k)$ of $\mathcal{L}$.\\
\noindent \textbf{Output:} The set $\varphi(\calS) =_{def}
\{(a_1,\dots,a_k)\in D^k: (\calS, a_1,\dots, a_k) \models
\varphi(x_1,\dots,x_k)\}$.

\medskip

In the following, query languages $\mathcal{L}$ are always
specified by fragments of first-order logic. The query $\calS
\mapsto \varphi(\calS)$ is often identified with the formula
$\varphi$ itself.

In this paper, we consider two different kinds of signature
$\sigma$: either $\sigma$ contains relation symbols only or it
contains relation and function symbols of arity at most one. In
the first case, $\sigma$ is said to be \textit{relational}, a
$\sigma$-structure will often be denoted by $\mathbf{db}$ and a
$\sigma$-query by $Q$. In the second case, $\sigma$ is said to be
\textit{ unary functional} or, for short, \textit{functional}, a
$\sigma$-structure is often denoted by $\calF$ and a
$\sigma$-query by $\varphi$.

By making  syntactic restrictions on the formula $\varphi$, one
may define a number of query problems. As we will see, the choice
of the kind of signature has some influence also and we will
define both relational query problems and  the associated
functional query problems. In what follows we briefly recall the
basics about "classical" conjunctive queries and revisit this
notion by introducing a new kind of functional query problem.

\subsection{Conjunctive queries}~\label{SEC Definitions of query problems}

 Conjunctive queries can
be seen as select-join-project queries (with renaming of
variables). Logically speaking, they are equivalent to queries
expressed by first-order relational formulas with existential
quantification and conjunction, i.e., of the form:
\[
Q(y_1,\dots,y_b) \equiv \exists x_1 \dots \exists x_a \
\phi(x_1,\dots,x_a,y_1,\dots,y_b)
\]

\noindent where $\phi$ is a conjunction of atoms over some
relational signature $\sigma$ and variables among $\tu x, \tu y$.
If $Q$ has no free variable the query is said to be \textit{Boolean}.

\begin{example}~\label{conjunctive queries} The two queries below are examples of conjunctive queries. 
\medskip

$
Q_1(y_1,y_2) \equiv \exists x_1 \exists x_2 \exists x_3 : \
R(x_1,y_1) \wedge S(x_1,y_2,x_2) \wedge T(y_2,x_3) \wedge
R(x_1, x_2) \\
$

$
Q_2 \equiv \exists x_1 \exists x_2 \exists x_3 \exists x_4 \exists x_5 : \
S(x_1,x_2,x_3) \wedge S(x_1,x_4,x_5) \wedge
R(x_3, x_5) \\
$

\noindent Query $Q_2$ is boolean.
\end{example}

\bigskip

An important and well-studied class of conjunctive queries are the
so-called \textit{acyclic} conjunctive queries. To each conjunctive
query $Q$ one associates the following hypergraph $\calH_Q =
(V,E)$ : its set of variables is $V=\textit{var}(Q)$  and its set of hyperedges is
$E= \{ \textit{var}(\alpha) : \alpha \  is\ an\ atom\ of\ Q \}$.
There exist various notions of acyclicity related to hypergraphs.
We have to use the most general one that is defined as follows. A hypergraph is
\textit{acyclic} if one can obtain the empty set by repeating the
following two rules (known as GYO rules, see~\cite{Graham-79,YuO-79}) until no change occurs:

\begin{enumerate}

\item Remove hyperedges contained in other hyperedges;

\item Remove vertices that appear in at most one hyperedge.

\end{enumerate}

\noindent As usual (see~\cite{Fagin-83}), a query is said to be
\textit{acyclic} if its associated hypergraph is acyclic. Denote
by $\ACQ{}$ the class of acyclic conjunctive queries.

\begin{figure}[t]
\begin{center}
\input{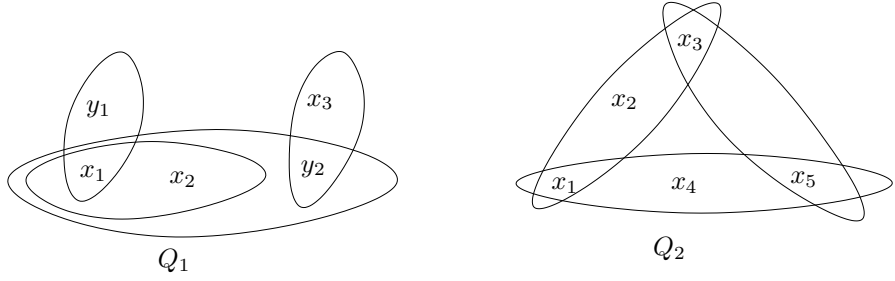}
\end{center}
\caption{The Hypergraphs of queries $Q_1$ and $Q_2$}~\label{example hypergraphs}
\end{figure}

\begin{example}
The hypergraphs associated to queries $Q_1$ and $Q_2$ of Example~\ref{conjunctive queries} are shown in Figure~\ref{example hypergraphs}. Applying GYO rules shows that $Q_1$ is acyclic and $Q_2$ is cyclic.
\end{example}

Conserving the same notion of acyclicity, one can enlarge this
class of queries by allowing inequalities between variables (as
defined in~\cite{PapadimitriouY-99}). This defines the larger class of so-called $\ACQineq{}$ queries.

\begin{example} Query $Q_3$ below is an example of an $\ACQineq{}$ query.

\[
\begin{array}{rl}
Q_3(y_1,y_2) \equiv \exists x_1 \exists x_2 \exists x_3 : & \
R(x_1,y_1) \wedge S(x_1,y_2,x_2) \wedge T(y_2,x_3) \wedge
R(x_1, x_2) \\
& \ \ \wedge y_1 \neq x_3 \wedge x_2 \neq x_1.
\end{array}
\]
\end{example}

\bigskip

Alternatively, it is well-known that a conjunctive query $Q(\tu
y)$ is acyclic  if and only if it has a \textit{join forest}
(called \textit{join tree} in case the forest is connected), that
is an acyclic graph $G_Q =(V,E)$ whose set of vertices $V$ is the
set of atoms of $Q$ and such that, for each variable $x$ that
occurs in $Q$, the set $A_x$ of relational atoms where $x$ occurs
is connected (is a subtree) in $G_Q$. Similarly, a conjunctive
query $Q(\tu y)$ with inequalities is in $\ACQineq{}$ if it has a
join forest $G_Q$. Note that $G_Q$ relies upon the relational
atoms but does not take into account the inequalities.

 One obtains another natural generalization of acyclic queries by allowing comparison atoms $x<y$.
 As proved by~\cite{PapadimitriouY-99} the evaluation problem of such queries is as difficult with respect to parameterized complexity as the clique problem (both are $W[1]$-complete problems) and hence is similarly conjectured to be f.p. intractable. Surprisingly, we will show that for the following class of acyclic queries with (restricted use of) comparisons, denoted by $\ACQplus{}$, the evaluation problem is exactly as difficult, with respect to time complexity, as that of $\ACQ{}$. A conjunctive query $Q$ with comparisons, i.e., atoms of the form $x \theta y$ where $x,y$ are variables and $\theta\in \{\ne, <, \leq, >, \geq \}$ is in  $\ACQplus{}$ if

 \begin{enumerate}

 \item it has a join forest $G_Q=(V,E)$ (defined as usual),

 \item for each comparison $x \theta y$  of $Q$, either $\{x,y\} \subseteq \textit{var}(\alpha)$, for some relational atom $\alpha$ of $Q$, or there is some edge $(\alpha,\beta)\in E$ in $G_Q$ such that $x\in \textit{var}(\alpha)$ and $y\in \textit{var}(\beta)$, and

 \item for each edge $(\alpha,\beta)\in E$, there is at most one comparison $x \theta y$ in $Q$ such that
 $x\in \textit{var}(\alpha)$ and $y\in \textit{var}(\beta)$.

\end{enumerate}

\noindent In other words, a conjunctive query with comparisons is
in  $\ACQplus{}$ if it has a join forest $G_Q$ and if each
comparison of $Q$ relates two variables inside the same vertex of
$G_Q$  or along an edge of $G_Q$, with globally at most one
comparison per edge. The reason for authorizing  only one comparison per edge of the tree will be explain later in Remark~\ref{REM more than one comparison}.

\begin{figure}[t]
\begin{center}
\input{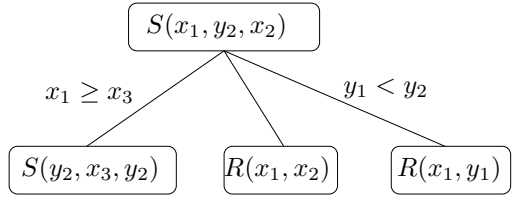}
\end{center}
\caption{Tree decomposition of query $Q_4$}~\label{join tree examples}
\end{figure}

\begin{example} Query $Q_4$ below is in $\ACQplus{}$. 

\[
\begin{array}{rl}
Q_4(y_1,y_2) \equiv \exists x_1 \exists x_2 \exists x_3 : & \
R(x_1,y_1) \wedge S(x_1,y_2,x_2) \wedge S(y_2,x_3,y_2) \wedge
R(x_1, x_2) \\
& \ \ \wedge y_1 < y_2 \wedge x_1 \ge x_3.
\end{array}
\]

\noindent Its join tree is shown in Figure~\ref{join tree examples}.
\end{example}

Finally, as defined in~\cite{FlumFG-02}, a query $Q(\tu y)$ is
said to be \textit{strict} if there exists a relational atom
$\alpha$ in $Q$ such that $\tu y \subseteq var(\alpha)$. We denote
by $\strictACQ{}$, $\strictACQplus{}$ and $\strictACQineq{}$ the
restrictions of the classes of queries $\ACQ{}$, $\ACQplus{}$,
$\ACQineq{}$, respectively, to strict queries.

\subsection{Conjunctive functional queries. }

In all this
part, $\sigma$ is a unary functional signature. In full generality, a conjunctive
functional query is a conjunctive query over some unary functional
signature $\sigma$. More precisely, it is of the form:

\[
\varphi(\tu y) \equiv \exists x_1 \dots \exists x_b : \
\bigwedge_{i=1}^h \tau_{i}(z_{i}) = \tau'_i(t_{i}) \wedge
\bigwedge_{i=1}^k U_i(\tau_i''(v_i))
\]

\noindent with $z_{i},t_{i}, v_{i}\in \var{\varphi}$, and $\tau, \tau',\tau''$  are terms made of compositions of unary function symbols of $\sigma$. For example, $\tau(x) = f_1f_2\dots f_k(x)$. Formulas are then interpreted on functional structures with \textit{totally} defined unary functions.

In this paper, formulas over a functional language are viewed as an analog of the well-knowm "tuple calculus". Then, for sake of clarity, we will adopt the following choices in the  presentation (these choices do not restrict the applicability of our results to queries of the most general form. See also Remark~\ref{REM composition disjunction}). In what follows, structures are considered as multisorted unary algebras i.e. as a collection of partially defined unary functions.
Let $\sg = \sg_{rel} \cup \sg_{fun}$ where $\sg_{rel}$ contains unary relation symbols only and $\sg_{fun}$ contains unary function symbols. A $\sg$-structure $\calF$ will verify :

\begin{itemize}

\item Its finite domain $D$ is such that $D$ is the union of all sets $T\in \sg_{rel}$. Also, for all $T_1,T_2 \in \sg_{rel}$, $T_1\cap T_2 = \emptyset$.

\item For each function $f\in \sg_{fun}$, there is a collection $T_{j_1},\dots,T_{j_k}$ of sets in $\sg_{rel}$, such that $f$ is defined over $\bigcup_{i\leq k} T_{j_i}$ (and undefined elsewhere) and has value in $D$.
   
\end{itemize}

This definition reflects the fact that each $T\in\sg_{rel} $ is seen as a set of tuples  with each function $f\in \sg_{fun}$ being a projection function from tuples to the domain $D$. The number of functions defined over $T$ is equal to the arity of the underlying relation that $T$ represents.

\bigskip

For what concerns $\sg$-formulas two restrictions will be adopted in this paper.

\begin{itemize}

\item Quantifications will always be relativized to some universe $X\in \sg_{rel}$ i.e. formulas are of the form $(\exists x \in T) \varphi$ which is equivalent to $\exists x \ T(x) \wedge \varphi$.

\item All atoms are of the form $x\in T$ for $T\in \sg_{rel}$ or $f(x)=g(y)$ for $f,g\in \sg_{fun}\cup \{Id\}$ where $Id$ is the identity function. Note that  composition of functions is not allowed here. 

\end{itemize}

\begin{example}
Formula $\varphi_1$ below defines a functional conjunctive query.

 \[
\begin{array}{rl}
\varphi_1(x) \equiv & \exists y \in T_1, \exists z \in T_2 : \\
& \qquad f_1(x) = g_1(y) \wedge f_2(x) = h_1(z) \wedge \\
& \qquad f_1(y) = g_2(z) \wedge f_1(z) \neq h_1(y) \wedge \\
& \qquad x \in T_1 
\end{array}
\]
\end{example}

 As in the relational setting, one can define a notion of
acyclic (unary) functional queries. The definition is even more
natural and simpler since it relies upon graphs instead of
hypergraphs.

\begin{definition}\label{DEF graphe requete}
Let $\varphi$ be a conjunctive functional query. The undirected
graph $G_{\varphi}= (V,E)$ associated to $\varphi$ is defined by:
$V=\var{\varphi}$ and for all distinct $x,y\in V$, $(x,y)\in E$
iff $\varphi$ contains at least one atom of the form $f(x) = g(y)$
for some $f,g\in \sg\cup \{Id\}$. The query $\varphi$ is
\textit{acyclic} if its graph $G_{\varphi}$ is acyclic.
\end{definition}

We denote by $\FACQ{}$ the class of acyclic (conjunctive)
functional queries. Again, one may authorize the use of negation
inside queries. We then denote by $\FACQineq{}$ the class of
acyclic functional queries $\varphi$ whose atoms are of one of the three forms
$f(x)=g(y)$, $f(x)\neq g(y)$, or $T(x)$, for $f,g \in \sigma \cup \{Id\}$ and $T \in \sigma$
(recall that the notion of acyclicity relies upon equalities only).

\begin{figure}[t]
\begin{center}
\input{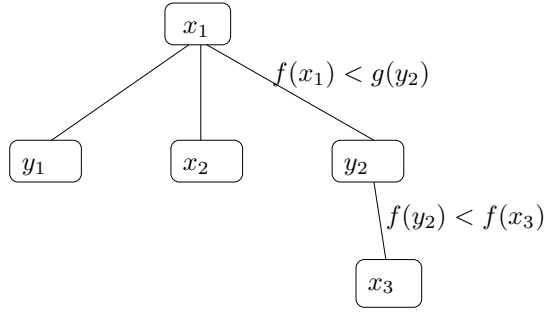}
\end{center}
\caption{Graph of queries $\varphi_2$ (without comparisons) and $\varphi_3$}~\label{functional query graph}
\end{figure}

\begin{example} The following query $\varphi_2$ belongs to $\FACQineq{}$.

\[
\begin{array}{rl}
\varphi_2(y_1,y_2) \equiv  & \exists x_1 \in T_1, \exists x_2 \in T_2, \exists x_3 \in T_2:
\\
& \ \ f(x_1)=f(y_1) \wedge g(x_1) = x_2 \wedge g(x_1) = f(y_2)
\wedge
g(y_2)=x_3 \wedge\\
& \ \ \wedge  x_3 \neq f(x_1) \wedge g(y_1) \neq f(y_2).
\end{array}
\]

The associated graph of  $\varphi_2$ is given in Figure~\ref{functional query graph}.
\end{example}

Similarly, let $\FACQplus{}$ denote the class of acyclic
functional queries $\varphi$ whose atoms are of the form
$f(x)=g(y)$ or $f(x)\theta g(y)$ or $U(x)$, for $f,g \in \sigma
\cup \{Id\}$, $U \in \sigma$, and $\theta\in \{\ne, <, \leq, >,
\geq \}$, whose associated graph $G_\varphi$ defined at Definition
\ref{DEF graphe requete}) is acyclic and for which the following
holds: if $f(x)\theta g(y)$ is a comparison atom of $Q$ for two
distinct variables $x$ and $y$ then $(x,y)\in E$ and, conversely,
for each edge $(x,y)\in E$, there is at most one comparison
$f(x)\theta g(y)$ in $Q$.

\begin{example}
Here is an example of $\FACQplus{}$ query.

\[
\begin{array}{rl}
\varphi_3(y_1,y_2) \equiv  & \exists x_1 \in T_1, \exists x_2 \in T_2 , \exists x_3 \in T_2 :
\\
& \ \ f(x_1)=f(y_1) \wedge g(x_1) = x_2 \wedge g(x_1) = f(y_2)
\wedge g(y_2)=x_3 \wedge\\
& \ \ \wedge  f(x_1) < g(y_2) \wedge f(y_2) \geq g(x_3).
\end{array}
\]

Its associated graph is given in Figure~\ref{functional query graph}.
\end{example}

In analogy with the notion defined above in the relational
setting, a functional query is said to be \textit{strict} if it
contains \textit{at most one free variable}. We denote by
$\strictFACQ{}$, $\strictFACQineq{}$ and $\strictFACQplus{}$ the
restrictions of the three above defined classes of queries to
strict queries. \raus{More generally, $\mathcal{C}_1$ denotes the
set of strict queries of any class $\mathcal{C}$ of functional
queries.}

\bigskip

In this paper, we will make extensive use of a class of queries
defined, roughly speaking,  as the complement of
acyclic functional queries. Let $\FFO{}$ be the class of first-order queries
defined by universal formulas in conjunctive normal form over some
 (unary) functional signature $\sg$, i.e., formulas of the form:

\[
\varphi(\tu y) \equiv \forall \tu x : \Et_{i\leq k} C_i(\tu x, \tu
y)
\]

\noindent where each $C_i$ is a  clause, i.e., a disjunction of literals of the form $(\neg)
f(z)=g(t)$ or $U(z)$ for $f,g \in \sigma \cup \{Id\}$ and
$U\in\sg$.

The negation of an $\FFO{}$ formula is clearly a disjunction of
conjunctive functional queries $\exists \tu x \ \neg C_i$. An
$\FFO{}$ query $\varphi$ is said to be \textit{acyclic} if
\text{each} query $\exists \tu x \ \neg C_i$ is acyclic. By
definition, the acyclicity of an $\FFO{}$ query can be read directly
on each clause of the query by looking at inequalities $f(z) \ne g(t)$
of the clause. The class of $\FFO{}$ acyclic queries is denoted by
$\FAFO{}$; its restriction to strict queries is obviously denoted by
$\strictFAFO{}$.

\begin{remark}~\label{REM composition disjunction}
In the formulas we consider, terms made of composition of functions are not autohrized at first sight. However, our results easily applies to this more general kind of formulas: for each term $\t (x) = f_1\dots f_k (x)$, one may add $\t$ as a new unary function sombol in the signature and pre-computes $\t (x)$, for eaxh $x\in D$, from $f_1$,\dots, $f_k$ in linear time. In this way, one can obtain an equivalent query problem but without composed terms.
Also, obviously, relativization and the use of partially defined functions do not play an essential role for what concerns the complexity results presented here.
\end{remark}

\subsection{Basic notions of complexity}

The model of computation used in this paper is the Random Access
Machine (RAM) with uniform cost measure (see~\cite{AhoHU-74,
GrandjeanS-02, GrandjeanO-04, FlumFG-02}). Basically, our inputs
are first-order structures and first-order formulas.

Let $E$ be a finite set or relation. We denote by $card(E)$ the
\textit{cardinality} of $E$. Let $[n]$ be the set $\{1,\dots,n\}$.
A set of cardinality $n$ is often identified with the set $[n]$.

The \textit{size} $|I|$ of an object $I$ is the number of
registers used to store $I$ in the RAM. If $E$ is the set $[n]$,
$|E|=card(E)=n$. If $R\subseteq D^k$ is a $k$-ary relation over
domain $D$, with $|D|=card(D)$, then $|R|=k.card(R)$: all the
tuples $(x_1,\dots,x_k)$ for which $R(x_1,\dots,x_k)$ holds must
be stored, each in a $k$-tuple of registers. Similarly, if $f$ is
a $k$-ary function from $D^k$ to $D$, all values
$f(x_1,\dots,x_k)$ must be stored and $|f|=|D|^k$.

 If $\varphi$ is a first-order formula, $|\varphi|$ is the number of
occurrences of variables, relation or function symbols and
syntactic symbols: $\exists, \forall, \wedge, \vee, \neg, =,  "(",
")", ","$. For example, if $\varphi \equiv \exists x \exists y \
R(x,y) \wedge \neg (x = y)$ then $|\varphi|=17$.

All the problems we consider in this paper are parameterized
problems: each takes as input a list of objects $I$ (e.g., a
$\sigma$-structure $\calS$ and a formula $\varphi$) together with
a parameter $k$ (e.g., the size of $\varphi$) and outputs an
object $S$ (e.g. the result of the query $\varphi(\calS)$).

A problem \textbf{P} is  computable in time
$f(k).T(|I|,|S|)$ for some function $f: \N \rightarrow \R^+$ if
there exists a RAM that computes \textbf{P} in time (i.e., the number
of instructions performed)  bounded by $f(k).T(|I|,|S|)$ using
space i.e., addresses and register contents also bounded by
$f(k).T(|I|,|S|)$~\footnote{This last restriction on addresses and
register contents forces the RAM to use its memory in a "compact"
way with space not greater than time.}. The notation
$O_k(T(|I|,|S|))$ is used when one does not want to make precise
the value of function $f$. 

\begin{definition}
Let $T$ be a polynomial function. A property $P$ is \textit{fixed-parameter tractable} if it is computable in time 
$f(k).T(|I|,|S|)$. When $T$ is of the form $T(n,p) = (n\times p)$, $P$ is said to be \textit{fixed-parameter linear}.
\end{definition}

It is easy to see that one obtains the same complexity measure if
instead of the uniform cost the logarithmic cost is adopted, i.e.,
if the time of each instruction is the number of bits of the
objects it manipulates. E.g., if the "uniform" time (and space)
complexity is $O_k(|I|,|S|)$ then the corresponding "logarithmic"
time complexity is $O_k(|I|.|S|.\log (|I|.|S|))$ which is at most
(and in fact less than) $O_k(|I|.\log |I|.|S|.\log |S|)=
O_k(|I|_{bit}.|S|_{bit})$ where $|I|_{bit} = \Theta(|I|.\log |I|)$
denotes the number of bits, i.e. the size in the logarithmic cost view, of the input $I$.

\section{Translating relational queries into functional queries}~\label{SEC From CQ to FQ}

The transformation of (acyclic) queries to be constructed in this
section is very similar to the translation of the domain
relational calculus into the tuple relational calculus in the classical framework of
database theory. Although one needs to examine carefully all the
details, the idea is very simple.

We want to transform each input $(\sg, Q, \textbf{db})$ of a
relational query problem into an input
$(\sg',\varphi_Q,\calF_{\textbf{db}})$ of a (unary) functional
query problem so that, among other things, $Q(\textbf{db})$ is
some projection of the relation  $\varphi_Q(\calF_{\textbf{db}})$.
Let us describe successively the transformation of the structure
and the corresponding transformation of the query.

\subsection{Transformation of the structure} 

Let $\textbf{db}$ be a
relational $\sg$-structure $\textbf{db}=\st{D}{\su{R}{q}}$ with each $R_i$ of arity $m_i$. For
convenience and simplicity (but w.l.o.g.), assume that there is no
isolated element in $D$, i.e., for each $x\in D$, there exists
$i\leq q$ and some tuple $t$ in $R_i$ to which $x$ belongs. Let $m = max_{i\leq q} m_i$ be the maximal arity among relations $R_i$s. The associated
functional $\sg'$-structure is defined as follows:

\[\calF_{\textbf{db}} = \st{D'}{D,\su{T}{q},\su{f}{m}}\]

\noindent where the domain $D'$ is the disjoint union of $q+1$
sets $D' = D \cup T_1 \cup \dots \cup T_q$ where $D$ is the domain
of $\textbf{db}$ and each $T_i$, $1\leq i \leq q$, is a set of
elements identified to the tuples of $R_i$
($card(T_i)=card(R_i)$); each of $D$, $T_1$, \dots, $T_q$ is a
unary relation of $\calF_{\textbf{db}}$; each $f_j$, $1\leq j \leq
m$, is a unary function.

Functions $f_j$ are defined as follows. For each $R_i$ of arity $m_i\leq m$ ($1\leq i \leq q$) and for each   $t\in T_i$ that represents the tuple $(e_1,\dots,e_{m_i})$ of $R_i$, set $f_1(t)=e_1, \dots, f_{m_i}(t)=e_{m_i}$. Intuitively, each $f_j$ is the $j^{th}$ projection for each tuple, it is obviously defined on sets $T_i$ that represents relations $R_i$ with $m_i \geq j$, else it is undefined.
Clearly, the functional structure $\calF_{db}$ encodes the whole
database structure $\textbf{db}$. We first have to prove the
following result.

\begin{proposition}
The transformation $\textbf{db} \mapsto \calF_{\textbf{db}}$ is
computable in linear time $O(|\textbf{db}|)$.
\raus{ ~\footnote{If the
$R_i$s are not of the same arity and considering the fact that we
are in the classical logic setting that \textit{does not} allow
partially defined functions, the precise bound of the reduction is
$O(\arity{\sg}.card(\textbf{db}))$  where $\arity{\sg}$ is the
maximal arity of the relational symbols of $\sg$ and
$card(\textbf{db})$ corresponds to the number of tuples of the
database (including the domain $D$). As $\sg$ is part of the
input, the reduction does not seem to be linear anymore. However,
this is not a major drawback:  all the complexity results that
will be presented in this paper are expressed in terms of the
domain size of the unary structure which corresponds to the number
of tuples of the original database. Also, it will be clear that
all the results remain true in the context of first-order logic on
partial unary functions, i.e., multi-sorted unary algebras seem to
be the most adapted framework}.
} 
\end{proposition}

\begin{proof}
Since the transformation is immediate, we only have to prove that
$|\calF_{\textbf{db}}| = O(|\textbf{db}|)$. It is essential to
notice that each $f_j$ is defined and described on some subset of $\bigcup_{i\leq q} T_i$ so that $|f_j| =   O(\sum_{i\leq q, j \leq m_i} |T_i|) =
O(\sum_{i\leq q, j \leq m_i} card(R_i))$ and hence $\sum_{j\leq m} |f_j| = O(
\sum_{i\leq q} m_i.card(R_i)) = O(\sum_{i\leq q} |R_i|) =
O(|\textbf{db}|)$. Finally, $\calF_{\textbf{db}} = |D'| + |D| +
\sum_{i\leq q} |T_i| + \sum_{j\leq m} |f_j| = O(|\textbf{db}|)$.
\end{proof}

\bigskip

\subsection{Transformation of the query}

The transformation is essentially the same for all the variants
($\ACQ{}, \ACQineq{}$, etc) of acyclic queries. We present it here
for $\strictACQineq{}$. Let $Q(y_1,\dots,y_b)$ denote an
$\strictACQineq{}$ query, i.e., a strict acyclic query with
inequalities of the form:

\[
Q(y_1,\dots,y_b) \equiv \exists x_1 \dots \exists x_a
\Psi(x_1,\dots, x_a,y_1,\dots,y_b)
\]

\noindent with $\Psi(\tu x, \tu y) \equiv \bigwedge_{1\leq i \leq
k } A_i \wedge I$ where the $A_i$'s are relational atoms and $I$
is a conjunction of variable inequalities $v\neq v'$ for
$v,v'\in\{\tu x,\tu y\}$.

By definition of the strict acyclicity, $Q$ has a join forest
$F=(V,E)$ whose set of vertices is $V=\{A_1,\dots,A_k\}$ so that
$\tu y \subseteq \var{A_1}$. We want to construct a conjunctive
functional $\sg'$-formula $\varphi_Q$  whose graph $G_{\varphi_Q}$
is exactly the acyclic graph $F$. Roughly, the idea is to replace
the $k$ atoms $A_1, \dots, A_k$ by $k$ variables $t_1,\dots, t_k$
that represent the corresponding tuples. For a relational atom
$A_u$, $1\leq u \leq k$, let $var_i(A_u)$ denote the $i^{th}$
variable of $A_u$: e.g., if $A_u$ is the atom $S(y_2,x_3,y_2)$
then $var_1(A_u) = var_3(A_u) = y_2$. As defined before, each
function $f_j$, $j\leq m$, of the functional structure
$\calF_{\textbf{db}}$ gives for each tuple $t$ (of a relation of
$\textbf{db}$) its $j^{th}$ field $f_j(t)$. E.g., the above
equality $var_1(A_u) = var_3(A_u)$ for $A_u= S(y_2,x_3,y_2)$ is
expressed by the formula $f_1(t_u) = f_3(t_u)$. The following
functional formula essentially mimics the description of formula
$Q$ and of its forest $F=(V,E)$:

\[
\varphi_Q (\tu t) \equiv \Psi_{rel}(\tu t) \wedge \Psi_{V}(\tu t)
\wedge \Psi_{E}(\tu t)
\wedge \Psi_{I}(\tu t).
\]

 Each conjunct of
$\varphi_Q$ is described precisely as follows.

\begin{itemize}

\item $\Psi_{rel}(\tu t)$ is $\bigwedge_{u\leq k}T_{v_u}(t_u)$ if
the atom $A_u$ is of the form $R_{v_u}(\ldots)$.

\item $\Psi_{V}(\tu t)$ is $\bigwedge_{u\leq k}\Psi_u$ where $\Psi_u$
is nonempty if $A_u$ has at least one repeated variable and
contains for each (repeated) variable that occurs at successive
indices $j_1,\dots,j_r$ of $A_u$ the conjunction
$\bigwedge_{i < r} f_{j_i}(t_u) = f_{j_{i+1}}(t_u)$.

\item $\Psi_{E}(\tu t)$ is $\bigwedge_{(A_u,A_v)\in E}\Psi_{u,v}$ where
$\Psi_{u,v}$ contains, for each variable $w$ that occurs both in
$A_u$ and $A_v$ with $(A_u,A_v)\in E$, one equality of the form
$f_{i}(t_u) = f_{j}(t_v)$ for two arbitrarily chosen indices $i,j$
such that $var_i(A_u) = w = var_j(A_v)$.

\raus{
\item $\Psi_{proj}(t_1,\tu y)$ contains for each free variable
$y_j$, $j\leq b$, one equality $f_i(t_1)=y_j$ for one arbitrarily
chosen index $i$ such that $var_i(A_1)=y_j$. }

\item  $\Psi_{I}(\tu t)$ is constructed as follows. For
each inequality $w\neq w'$ of $I$, choose (arbitrarily, again) two
atoms $A_u$ and $A_v$ so that $w$ (resp. $w'$) occurs in $A_u$
(resp. $A_v$) at index $i$ (resp. $j$). Replace $w\neq w'$ by the
inequality $f_i(t_u) \neq f_j(t_v)$. Let $\Psi_{I}$ be the
conjunction of all those inequalities.
\end{itemize}

Due to formula $\Psi_{rel}(\tu t)$, each quantified variable is relativized to some domain $T_i$.

\begin{example}
The following query:

\[
Q(y_1,y_2) \equiv \exists x_1 \exists x_2 \exists x_3 :
R_1(x_1,y_1,y_2) \wedge R_2(x_2,x_1,x_2) \wedge R_1(x_2, x_2, x_3
) \wedge y_1 \neq x_2,
\]

\noindent is translated into the formula $\varphi_Q(\tu t)$, with
$\tu t = (t_1,t_2,t_3)$, that is the conjunction of the following
formulas:

 \[
\begin{array}{rl}
\Psi_{rel}(\tu t) \equiv & T_1(t_1) \wedge T_2(t_2) \wedge T_1(t_3)\\
 \Psi_{V}(\tu t) \equiv  & f_1(t_2) =
f_3(t_2) \wedge f_1(t_3) =
f_2(t_3)\\
\Psi_{E}(\tu t) \equiv & f_1(t_1) = f_2(t_2) \wedge f_1(t_2) =
f_1(t_3) \\
\Psi_{I}(\tu t) \equiv & f_2(t_1) \neq f_1(t_2)\\
\end{array}
 \]

 Finally, it is easy to check that the following equality holds:

 \[
Q(\textbf{db}) = \{ (f_2(t_1),f_3(t_1)) : \ t_1\in D' \mbox{ and }
(\calF_{\textbf{db}},t_1) \models \exists t_2 \exists t_3
\varphi_Q(\tu t)\}.
 \]

 \noindent In other words, $Q(\textbf{db})$ is the result of the projection
 $\tu t \mapsto (f_2(t_1),f_3(t_1))$ applied to the relation
 $\varphi_Q(\calF_{\textbf{db}})$. Obviously, formula $ \exists t_2 \exists t_3
\varphi_Q(\tu t)$ is equivalent to the relativized formula:

\[
\exists t_1 \in T_1, \exists t_2 \in T_2 : \ T_1(t_3) \wedge \Psi_{V}(\tu t) \wedge \Psi_{E}(\tu t) \wedge \Psi_{I}(\tu t).
\] 
\end{example}

More generally, the transformation process described before yields
the following properties.

\begin{lemma}
Let $Q(y_1,\dots,y_b)$ be a query in $\strictACQineq{}$ (resp.
$\ACQineq{}$, $\strictACQ{}$, $\ACQ{}$, $\strictACQplus{}$,
$\ACQplus{}$). The following properties hold:

\begin{enumerate}
\item $\varphi_Q \in \strictFACQineq{}$ (resp.  $\FACQineq{}$,
$\strictFACQ{}$, $\FACQ{}$, $\strictFACQplus{}$, $\FACQplus{}$).

\item For each relational $\sg$-structure $\textbf{db}$,
the result $Q(\textbf{db})$ (of query $Q$ over $\textbf{db}$) is
obtained by some "projection" of the relation
$\varphi_Q(\calF_{\textbf{db}})$. More precisely, there are two
lists of indices $\su{i}{b}$ and $\su{j}{b}$ such
that~\footnote{In case $Q(\tu y)$ is a strict query with $\tu y
\subseteq var(A_1)$ then $j_1=\dots = j_b =1$.}

\[
\begin{array}{rl}
Q(\textbf{db}) = & \{ (f_{i_1}(t_{j_1}), \dots,f_{i_b}(t_{j_b})):
\\
& \mbox{ there exist } t_1,\dots, t_k \in D' \mbox{ such that }
(\calF_{\textbf{db}},\tu t) \models \varphi_Q(\tu t)\}.
\end{array}
 \]

 \noindent where $y_h = var_{i_h}(A_{j_h})$ for $h=1,\dots, b$.

\item $\vert \varphi_Q \vert = O(|Q|)$.
\end{enumerate}
\end{lemma}

\begin{proof} For simplicity of notation, let us still assume that
$Q$ belongs to $\strictACQineq{}$.

\begin{enumerate}
\item By construction, the graph $G_{\varphi_Q}$ is (up to
isomorphism) the join forest $F$ (associated to $Q$); this
corresponds to the conjunct $\Psi_E$. See also Remark~\ref{REM GYO rules}

\item By definition of the join forest $F$, the set of atoms where any fixed variable
of $Q$ occurs is connected in $F$. This implies that the conjunct
$\Psi_{V} \wedge \Psi_E$ exactly expresses which variables the
relational atoms $A_1,\dots, A_k$ of $Q$ share. Moreover, $\Psi_I$
correctly expresses the conjunction $I$ of inequalities of $Q$.
This proves that for each relational $\sg$-structure $\textbf{db}
= \st{D}{\su{R}{q}}$ where $\calF_{\textbf{db}} =
\st{D'}{D,\su{T}{q},\su{f}{m}}$ and for all $\tu y \in D^b$, it
holds:

\[
\begin{array}{c}
(\textbf{db},\tu y) \models Q(\tu y) \mbox{ \textit{iff} } \\
\mbox { there exists } t_1,\dots, t_k \in D' \mbox{ such that } \\
(\calF_{\textbf{db}},\tu t) \models \varphi_Q(\tu t) \mbox { and }
f_{i_h}(t_1) = y_h \mbox{ for each } h=1,\dots,b.
\end{array}
\]

\item Let $NbOcc$ denote the number of occurrences of variables
in $Q$. It is easy to see that:

\[
|\Psi_{V}| + |\Psi_E| = O(NbOcc).
\]

Clearly, we also have $|\Psi_I| = O(|I|)$ and $|\Psi_{rel}| =
O(k)$. That implies $|\varphi_Q| = O(|Q|)$.
\end{enumerate}

\end{proof}

\begin{remark}~\label{REM GYO rules}
Having a  join forest $F$ for $Q$ is not necessary to construct the acyclic formula $\Psi_E$ in $\varphi_Q$. There is an alternative way to obtain an equivalent $\Psi_E$ using the GYO rules (\cite{Graham-79,YuO-79}). Let $\calH_Q$ be the hypergraph associated to query $Q$. For each application of rule~$2$ (\textit{"remove vertices that appear in at most one hyperedge"})  nothing as to be done. However, each time  rule~$1$ (\textit{"remove hyperedges contained in other hyperedge"}) is applied to some atom $A_u$ and $A_v$, one proceed as for the original construction of $\Psi_E$: for each variable $w$ that occurs in $A_u$ and $A_v$, one equality of the form
$f_{i}(t_u) = f_{j}(t_v)$ for two arbitrarily chosen indices $i,j$
such that $var_i(A_u) = w = var_j(A_v)$ is constructed.

Applying these rules till $\calH$ is empty will result in a new acyclic formula $\Psi_E$. 
\end{remark}

\section{Samples of unary functions}~\label{SEC Samples of unary functions}

In this section, some simple combinatorial notions about unary
functions are defined. They may be seen as  some kind of set/table covering  problem. They will be essential for proving the main
results of this paper.


\begin{definition}\label{DEF echantillonnage}
Let $E,F$ be two finite sets and $\tu g=(g_1,\dots,g_k)$ be  a
tuple of unary functions from $E$ to $F$.  Let $P\subseteq [k]$
and $(c_i)_{i\in P}$ be a family of elements of $F$.
$(P,(c_i)_{i\in P})$ is said to be a {\em sample} of $\tu g$
(indexed by $P$) over $E$ if
$$
E=\bigcup_{i\in P}g_i^{-1}(c_i).
$$
\noindent where $g_i^{-1}(c_i)$ is the set of preimages of $c_i$
by function $g_i$. A sample is said to be {\em minimal} if, moreover, for all
$j\in P$:
$$
E\neq\bigcup_{i\in P\setminus\{j\}}g_i^{-1}(c_i).
$$
Finally, if $(P,(c_i)_{i\in P})$ is a sample (resp. minimal
sample) of $\tu g$ over $E$, the family of sets $(g_i^{-1}(c_i))_{i\in P}$ is called a
{\em covering} (resp. {\em minimal covering}) of $E$ by $\tu g$.

Samples will often simply be denoted $(c_1,\dots,c_k)$ with $c_i= \ '-'$ when $i\not\in P$.
\end{definition}

\bigskip

\begin{example}
Let $\tu g = (g_1,g_2,g_3)$ be the following tuple of unary functions over some domain/table $T$  with tuples $a,b,c,d,e$.

\[
\begin{array}{c|c|c|c|}\cline{2-4}
 & g_1 & g_2 & g_3 \\ \cline{2-4}
a & 1 & 2 & 4 \\ 
b & 1 & 5 & 1 \\
c & 3 & 2 & 4 \\
d & 3 & 5 & 3 \\
e & 5 & 2 & 4 \\\cline{2-4}
\end{array}
\]

It is easily seen that the tuples $(1,2,3),(1,5,4),(3,2,1)$ and $(-,5,4)$ are the samples of $\tu g$ over $T$. Among them, $(1,2,3),(3,2,1)$ and $(-,5,4)$ are minimal.
\end{example}

\bigskip

\begin{remark}\label{REM minimal sample}
Let $P \subseteq [k]$ and $(c_i)_{i\in P}$ be a sample of $\tu g$
over $E$. Then, there exists $P'\subseteq P$ such that
$(c_i)_{i\in P'}$ is a minimal sample of $\tu g$ over $E$.
Informally, it is obtained by repeating the following steps as
long as possible:

- pick a $j$ from $P$ such that $E=\bigcup_{i\in
P\setminus\{j\}}g_i^{-1}(c_i)$

- set $P\leftarrow P\setminus\{j\}$.

\medskip

\noindent Note that the only minimal sample of $\tu g$ over the
empty set is $(-,-,\dots,-)$
\end{remark}

\bigskip

In the rest of this section, problems about minimal samples are
defined and their complexities are studied. Those problems will
play a key role in the paper.

\bigskip

\noindent \pb{Min-Samples}\\
\noindent \begin{tabular}{rl}
\textbf{Input:} & \parbox[t]{300 pt}{two finite sets $E$ and $F$ and a $k$-tuple of unary functions $\tu g =(\su{g}{k})$ from $E$ to $F$.}\\
\textbf{Parameter:} &  integer $k$.\\
\textbf{Output:} & \parbox[t]{300 pt}{the set of minimal samples of $\tu g$ over $E$.}\\
\end{tabular}

\bigskip


\begin{lemma}\label{RES borne sup} Let $E$, $F$, $\tu g$ be an  input of \pb{Min-Samples}.

\begin{enumerate}
\item There are at most $k!$ minimal samples of $\tu g$ over $E$.

\item Problem \pb{Min-Samples} can be solved in time $O_k(|E|)$.
\end{enumerate}

\end{lemma}

\begin{proof}
Let us identify $E$ with the set $\{1,\dots,n\}$. We describe the
construction of a tree $T$ of depth $n$ with at most $k!$ leaves and
hence at most $k!|E|=O_k(n)$ nodes. The leaves represent all the
minimal samples of $\tu g$ over $E$. Level~$i$ of the tree corresponds to
element~$i$ of $E$. Each node $x$ of level~$i$ is labelled by a subset
$P^x\subseteq [k]$ and by a sample $(c_j^x)_{j\in P^x}$ of $\tu g$ over $\{1,\dots,i\}$ .
The root $r$ of the tree is labelled by $P^r=\emptyset$.

Let $i=1$. There are at most $k$ possibilities to cover
element~$1$ with $g_h(1) = c_h$ $(h=1,\dots,k)$. Then, the root of
$T$ has $k$ children $x_h$ each labelled by
$(P^{x_h}=\{h\},c_h)$.

At each level~$i$ ($i=2,\dots,n$), the same strategy is used. Let
$x$ be a node of level~$i-1$ labelled by $(P^x,(c_j^x)_{j\in
P^x})$. The set of children $y$ of $x$ labelled by
$(P^y,(c_j^y)_{j\in P^y})$ will correspond to all the
possibilities to extend the covering of $\{1,\dots,i-1\}$ by
$(P^x,(c_j^x)_{j\in P^x})$ in a minimal way in order to cover node
$i$ (if $i$ is not already covered).

Testing whether $i$ is already covered, i.e., if $i \in \bigcup_{j\in
P^x}g_j^{-1}(c_j^x)$ can be done in constant time $O_k(1)$: it suffices to
test the disjunction $\bigvee_{j\in P^x}g_j(i)=c_j^x$. Two cases
may occur:

\begin{itemize}
\item
Either $i \in \bigcup_{j\in P^x}g_j^{-1}(c_j^x)$. In this case,
node $x$ has a unique child node $y$ of level~$i$ with $P^{y}=P^x$
and $\forall j \in P^x$, $c_j^y=c_j^x$.

\item
Or $i\not \in \bigcup_{j\in P^x}g_j^{-1}(c_j^x)$. Two subcases may
hold.

\begin{itemize}

\item
Either $P^x=[k]$. Then, it is not possible to cover element~$i$,
the construction fails and stops here for that branch.

\item
Or $P^x\varsubsetneq [k]$. Then, for each $h\in [k]\backslash
P^x$, one constructs a child node $y$ for $x$ such that:
$P^{y}=P^x\cup\{h\}$, $c_j^y = c_j^x$ for $j\in P^x$ and
$c_h^y=g_h(i)$. Node $y$ and its label can be constructed in
constant time.
\end{itemize}
\end{itemize}

That process ends after $O_k(|E|)$ steps with a tree of size
$O_k(|E|)$ whose leaves represent the (up to) $k!$ minimal samples
of $\tu g$ over $E$: $(c_1,\dots,c_k)$ with $c_i= \ '-'$ when $i\not\in P$.~\footnote{To be completely rigorous, we should
mention that some of the (up to) $k!$ samples that our algorithm
constructs may be not minimal. The essential property is that, by
construction, each minimal sample (of $\tu g$ over $E$) is
included in (at least) one of the constructed samples. The
algorithm above should be completed by a variant of the algorithm
of Remark~\ref{REM minimal sample} that extracts from a sample all
the minimal samples that it contains. Note that the whole
additional time required is $O_k(1)$ and that some minimal samples
may be repeated.}.
\end{proof}

\bigskip

We will also need a more elaborate problem about samples. Let $l$
be an integer and $v$ be a function from $E$ to $E^l$. The image
set of $v$ is denoted by $v(E)$. It is clear that the collection
of sets $v^{-1}(\tu a)$ for $\tu a =(a_1,\dots,a_l)\in
v(E)\subseteq E^l$ forms a partition of $E$. Let us define the
following problem.

\bigskip

\noindent \pb{Min-Samples-Partition}\\
\noindent \begin{tabular}{rl}
\textbf{Input:} & \parbox[t]{300 pt}{two finite sets $E$ and $F$, two
integers $k$ and $l$, a $k$-tuple of unary functions
$\tu g =(\su{g}{k})$ from $E$ to $F$ and a function $v$ from $E$ to $E^l$.}\\
\textbf{Parameter:} &  integers $k$ and $l$.\\
\textbf{Output:} & \parbox[t]{300 pt}{for each $\tu a \in v(E)\subseteq E^l$, the set $M(\tu a)$ of minimal samples of $\tu g$ over $v^{-1}(\tu a)$.}\\
\end{tabular}

\bigskip


\begin{lemma}\label{RES borne sup 2}
Problem \pb{Min-Samples-Partition} can be solved in time
$O_{k,l}(|E|)$.
\end{lemma}

\begin{proof} The algorithm is the following.
\begin{enumerate}

\item~\label{RES borne sup 2: step 1} Compute the set $S= \{(v(x),x): \ x\in E\}$ in time $O_l(|E|)$.

\item~\label{RES borne sup 2: step 2} Sort $S$ by values of $v(x)$: this computes the
partition $(v^{-1}(\tu a))_{\tu a\in v(E)}$ of $E$ in time
$O_l(|E|)$.

\item~\label{RES borne sup 2: step 3} For each $\tu a\in v(E)$, compute the set $M(\tu a)$ of minimal samples of $\tu g$ over $v^{-1}(\tu a)$  in time $O_k(|v^{-1}(\tu a)|)$ (by Lemma~\ref{RES borne
sup}). The total time required for this last step is
$O_k(\sum_{\tu a\in v(E)}|v^{-1}(\tu a)|)=O_k(|E|)$.
\end{enumerate}
\end{proof}

\bigskip

\begin{remark}
The "sampling" problems  and their algorithms that are involved
in Lemmas~\ref{RES borne sup} and~\ref{RES borne sup 2} can be seen as generalizations of the well-known k-VERTEX COVER
problem in graphs and its algorithm of parameterized  linear complexity
$O_k(|G|)$ (see~\cite{DowneyF-99}). 
 
Let $G=\st{V}{E}$ be a graph and $\calF_G=\st{D}{f_1,f_2}$ be its functional representation : $D=V\cup E$ and for each $e\in E$, $f_1(e)$ and $f_2(e)$ describe the endpoints of $e$. Then $C=\{c_1,\dots,c_k\}\subseteq V$ is a k-VERTEX COVER of  
$G$ if:

\[
\forall x\in E, f_1(x)=c_1 \vee \dots \vee f_1(x)=c_k \vee f_2(x)=c_1 \vee \dots \vee f_2(x)=c_k.
\]

\noindent In other words, $(c_1,\dots,c_k,c_1,\dots,c_k)$ is a $(f_1,\dots,f_1,f_2,\dots,f_2)$-sampling of $V$.
\end{remark}

\section{The complexity of functional acyclic queries}
~\label{SEC The complexity of generalized  acyclic queries}

Roughly, the main technic of this paper shows among other things
that it is possible to eliminate quantified variables in an
acyclic conjunctive query (by transforming both the query and the
structure)without overhead in the query evaluation process, i.e.,
so that evaluating the query so simplified is just as hard as
evaluating the original query.

For the sake of clarity, the main result will first be stated in
the context of $\FAFO{}$ queries. Let us explain the method on a
very simple example. Let $\varphi$ be the following  Boolean $\FAFO{}$ query (without
negation and only two variables):

\[
\varphi \equiv \forall x \forall y : f_1(x) = g_1(y) \vee \ldots
\vee f_k(x) = g_k(y)
\]

A first naive approach for evaluating $\varphi$ against a given
unary functional structure $\calF = \st{D}{\tu f, \tu g}$ consists
in testing the truth value of the matrix for any possible value of
$(x,y)$: that requires a time $O_k(|D|^2)$. Alternatively,
$\varphi$ can be interpreted as follows: for each value of $x$,
the family of sets $g_i^{-1}(f_i(x))$, for $i\in \{1,\dots,k\}$,
is a covering of $D$. In other words, for each $x$, there exists a
sample $(P,(c_i)_{i\in P})$ of $\tu g$ over $D$ (with initially
$P=[ k]$) that "agrees" with values of $\tu f(x)$, i.e., such
that:

\begin{equation}\label{propriete min sample}
\bigwedge_{i\in P} f_i(x) = c_i \mbox{ holds. }
\end{equation}

\noindent Such a sample can be chosen among minimal ones (recall
Remark~\ref{REM minimal sample}). Then, evaluating $\varphi$
against $\calF$ can be done as follows. First, the set of the (up
to) $k!$ minimal samples of $\tu g$ over $D$ is computed. Then,
for each $x$, it is looked for one of these minimal samples that
satisfy Property~\ref{propriete min sample}. Because of
Lemma~\ref{RES borne sup}, the whole process requires $O_k(|D|)$
steps.

With some more work, this basic idea can be  extended to the
general case of (non necessarily Boolean) acyclic queries where
both equalities and inequalities are allowed. This is achieved through the main result
that follows.

\begin{theorem}~\label{RES: FO acyclic}
The $\strictFAFO{}$ query problem can be solved in time
$f(|\varphi|).|D|$ where $D$ is the domain of the input structure
$\calF$, $\varphi$ is the input formula and $f$ is a fixed
function from $\mathbb{N}$ to $\mathbb{R}^+$.
\end{theorem}

\begin{proof}
Let  a unary functional structure $\calF$ of domain $D$ and a
formula $\varphi(x) \equiv \forall \tu y \phi(x, \tu y)$ with $\tu
y = (y_1,\dots,y_d)$ be the inputs of an $\strictFAFO{}$ query
problem. Since clauses may be reduced independently, it can be
supposed that $\varphi(x)$ contains only one clause $\phi$. The proof is
done by induction on the number of variables of $\varphi$.

Let $G_{\varphi}$ denote the acyclic graph associated to $\varphi$
whose set of vertices is $\textit{var}(\varphi)$. Recall that
$G_{\varphi}$ only takes into account the \textit{inequalities} of
the clause $\phi$. Without loss of generality, assume that
$G_{\varphi}$ is connected, i.e., is a tree $T$ and choose $x$ as
the root of $T$. Order the nodes of $T$, i.e., the variables of
$\varphi$, by increasing levels from the root to the leaves, as
$y_0=x, y_1, \dots, y_d$. Note that the restriction of $T$ to the
subset of variables $\{y_0,y_1,\dots, y_i\}$ for $i\leq d$ is a
subtree $T_i$ where $y_i$ is a leaf.  The variables of $\varphi$
except $y_0=x$ will be eliminated one by one according to this
ordering. Let $y_{i_0}$, $i_0\leq d-1$, be the parent of leaf
$y_d$ in $T$.   W.l.o.g., assume  that $\varphi$ is of the
form~\footnote{In case $\varphi$ contains one-variable atoms of
the form $(\neg) u(y_d)=v(y_d)$ or $(\neg) U(y_d)$, we can easily
replace them by two-variable positive atoms by expanding the
signature and the structure by new unary functions computable in
linear time}:

\[\varphi(y_0) \equiv \fa y_1 \dots  \fa y_d :  \Ou_{j\leq l} v_j(y_d)\neq u_j(y_{i_0}) \vee
\psi (\tu y) \ou \Ou_{j\leq k} g_j(y_d) = f_j(y_{p_j})
\]

\noindent where $\tu y$ is now the $d$-tuple of variables $(y_0, y_1,
\dots,y_{d-1})$ , the $u_j$, $v_j$, for $1\leq j \leq
l$, and $f_j$, $g_j$, for $1\leq j \leq k$ are unary function
symbols, $\psi(\tu y)$ is an acyclic clause over $\tu y$ of
associated graph $G_{\psi} = T_{d-1}$, and for each $j\leq k$, $0
\leq  p_j \leq d-1$. Replacing our disjunction of negated atoms by
an implication, one obtains:

\[\varphi(y_{0}) \equiv \fa y_1 \dots \fa y_{d-1} \fa y_d :  \tu v(y_d)=\tu u(y_{i_0})\imp
(\psi (\tu y)\ou\Ou_{j\leq k} g_j(y_d) = f_j(y_{p_j}))
\]

\noindent  where $\tu v(y_d)=\tu u(y_{i_0})$ stands for $
\Et_{j\leq l} v_j(y_d) =  u_j(y_{i_0})$. Formula $\varphi$ can be
equivalently written as:

\[
\varphi(y_0) \equiv \fa y_1 \dots \fa y_{d-1} :  \psi (\tu y) \ou
[ \fa y_d \in \tu v^{-1}(\tu u(y_{i_0}))  \Ou_{j\leq k} g_j(y_d) =
f_j(y_{p_j})].
\]

The second  disjunct states that $(f_j(y_{p_j}))_{j\leq k}$
is a sample of $\tu g$ over $\tu v^{-1}(\tu u(y_{i_0}))$
and hence contains such a minimal sample.

 The family $M = \{ (b,M(\tu u (b))): b\in D\}$ of the sets of minimal
 samples $M(\tu u (b)) = \{(c^h_1(b),\dots,c^h_k(b)): \ 1\leq h \leq k!\}$  of $\tu g$ over $\tu v^{-1}(\tu u(b))$ is computed by Algorithm~A below (since the number of minimal samples is only bounded by $k!$, there may be repetitions of identical samples).

\bigskip

\noindent \textbf{Algorithm A:}

\begin{enumerate}

\item
Compute the family $A= \{(\tu a, M(\tu a)):\ \tu a \in \tu v(D)
\}$ where $M(\tu a)$ is the set of minimal samples of $\tu g$ over
$\tu v^{-1}(\tu a)$: this amounts to solve the problem
\pb{Min-Samples-Partition} in time $O_{k,l}(|D|)$ by
Lemma~\ref{RES borne sup 2} for $E= F= D$ and $v = \tu v$.

\item Sort $A$ in lexicographic order according to $\tu a$.

\item Compute and lexicographically sort the set $B= \{(\tu u (b),b): \ b\in D \}$ in time $O_{l}(|D|)$.

\item Merge the sorted lists $A$ and $B$ (in time  $O_{k,l}(|D|)$) to compute the set
$$C= \{(\tu u (b), M( \tu u (b)),b): \  b\in D \}$$

\raus{
\noindent (Note that , by definition, $M( \tu u
(b))=\emptyset \}$ if $\tu u (b)\not \in \tu v(D)$
because in this case the set to be covered $\tu v^{-1}(\tu u(b))$
is empty. ) 
} 

\item Return the family of sets (of minimal samples) $M=\{ (b, M( \tu u (b))): \  b\in D \}$

\end{enumerate}

Hence, the time complexity of Algorithm~A is $O_{k,l}(|D|)$. 
In the set $M$, we are interested, of course, by the elements $b$ for which $M( \tu u (b))$ is not empty. Let: 

\[
K = \{b: \ M( \tu u (b)) \neq \emptyset\}.
\]

We
now eliminate variable $y_d$ by expanding the signature of the
query and the structure (a classical method in quantifier
elimination). New unary relations $K$, $S_j^h$ and functions $c_j^h$
(for $h\leq k!$ and $j\leq k$) are introduced.

Functions $c_j^h$ are those that appear in the description of the
minimal samples.  Predicates $S_j^h$ are defined from $P^h$ as
follows: for all $j,h$ and $y\in D$, 

\[S_j^h(y) \ssi y\in K \mbox{ and } c_j^h \neq \ '-'.\]

 Let $\calF'$ be the expansion of structure
$\calF$ defined as $\calF' = (\calF,(S^h_j,c^h_j)_{h\leq k! ,
j\leq k})$. Let now $\varphi'$ be the following formula having $d$
variables $y_0,\dots,y_{d-1}$ (recall $i_0<d$):

\[
\varphi'(y_{0}) \equiv \fa y_1 \dots \fa y_{d-1} : \psi (\tu y)
\ou [ K(y_{i_0}) \et \Ou_{h\leq k!}\Et_{j\leq k} (S^h_j(y_{i_0}) \imp f_j(y_{p_j})
= c_j^h(y_{i_0}))]
\]

\noindent The last part of formula $\varphi'$ simply asserts that
if $j$ belongs to the index set of the $h$-th minimal sample of
$\tu g$ over $\tu v^{-1}(\tu u(y_{i_0}))$ then, $f_j(y_{p_j})$
must be equal to the $j$-th value of the $h$-th minimal sample.
That means that $(f_j(y_{p_j}))_{j\leq k}$ contains a  minimal
sample of $\tu g$ over $\tu v^{-1}(\tu u(y_{i_0}))$. It is then
clear that, for each possible value $a$ of $x=y_{0}$, it holds
$(\calF,a) \models \varphi(x) \ssi (\calF',a) \models
\varphi'(x)$, that means $\varphi(\calF)=\varphi'(\calF')$. Notice
that the last part of formula $\varphi'$ does not really introduce
negative atoms: it can be rephrased as $ \Ou_{h\leq k!}\Et_{j\leq
k} (S^h_j(y_{i_0}) = 0 \vee f_j(y_{p_j}) = c_j^h(y_{i_0}))$ where
$S^h_j$ is now regarded as a unary function from $D$ to $\{0,1\}$.
From the previous paragraphs, the two following facts also clearly
hold.

\begin{fact} The expansion $\calF'$ of structure $\calF$ can
be computed in time $O_{k,l}(|D|)$.
\end{fact}

\begin{fact} Formula $\varphi'(x)$ can be easily transformed
into a conjunction of acyclic clauses each having $d$ variables
and associated tree $T_{d-1}$.
\end{fact}

By iterating this process $d$ times, i.e., eliminating
successively variables $y_d,y_{d-1},\dots,y_1$, one obtains in
time $O_{\varphi}(|D|)$ an expansion $\calF'$ of $\calF$ and a
quantifier-free formula $\varphi'(x)$ with only one variable
$x=y_0$. It is clear that the final query $\varphi'(\calF')=\{a\in
D: (\calF',a) \models \varphi'(x) \}=\varphi(\calF)$ can be
computed in linear time $O(|\varphi'|.|D|)$.
\end{proof}

\begin{remark}[On the constant value  $f(|\varphi|)$]~\label{REM constant f}
In the worst case, the value of $f(|\varphi|)$ may be huge: each
elimination step may introduce a number of new atoms bounded by
$k!$ (and requires to put the new formula in conjunctive normal
form for the next step).

A very interesting particular case concerns
$\strictFAFO{}$-queries without positive atoms (closely related to
the $\strictACQ{}$ problem). In that case, formula $\varphi(x)$ is
of the following form, for some $i_0\leq d$:

\[
\begin{array}{rl}
\varphi(x) & \equiv \fa y_1 \dots \fa y_{d-1} \fa y_d : \Ou_{j\leq
l}
v_j(y_d)\neq u_j(y_{i_0}) \vee \psi (\tu y) \\
 & \equiv \fa y_1 \dots \fa
y_{d-1}  : \psi (\tu y) \ou \neg \exists y_d  (\tu v(y_d)=\tu
u(y_{i_0}))
\end{array}
\]

\noindent with $\tu y = (y_0,\dots,y_{d-1})$. It is easy to see
that one can compute, in time $O(l.|D|)$, the set $D_0$ of
elements $y\in D$ such that $\tu u(y)\in \tu v(D)$ (i.e., such
that there exists $y_d$ with $\tu v(y_d)=\tu u(y)$). By enlarging
the signature, the formula $\varphi$ can be transformed into an
equivalent formula without variable $y_d$ (also denoted by
$\varphi$ for convenience):

\[
\varphi(x) \equiv \fa y_1 \dots \fa y_{d-1}  : \psi (\tu y) \ou
\neg D_0(y_{i_0})
\]

\noindent Note that, although a new atom $D_0(y_{i_0})$ has been
introduced, the sum of the number of quantifiers plus the number
of literals of $\varphi$ has been decreased by $l$.  Summing up
the costs of all the steps, it yields that $\strictFAFO{}$-queries
without positive atoms can be evaluated in time
$O(|\varphi|.|D|)$.
\end{remark}

\bigskip

We are now able to state the consequences of our results, first in
the context of acyclic  conjunctive functional queries.

\begin{theorem}~\label{RES strictFACQineq and strictFACQ}
The query problem $\strictFACQineq{}$ (resp. $\strictFACQ{}$)
can be solved in time $f(|\varphi|).|D|$ (resp.
$O(|\varphi|.|D|)$).
\end{theorem}

\begin{proof}
Let $\calF$ and $\varphi(x)$ be inputs of the $\strictFACQineq{}$
(resp. $\strictFACQ{}$) problem. By definition, $\neg \varphi(x)$
defines the $\strictFAFO{}$ query (resp. $\strictFAFO{}$ query
without positive atoms) whose output is $D\setminus
\varphi(\calF)$. By Theorem~\ref{RES: FO acyclic}, Remark~\ref{REM
constant f} and the fact that $\varphi(\calF)$ can be computed
from $D\setminus \varphi(\calF)$ is time $O(|D|)$, we are done.
\raus{ The algorithm that solves the  $\strictFACQplus{}$ problem
in time $O(|\varphi|.|D|)$ is quite similar to the algorithm given
for $\strictFACQ{}$ (see Remark~\ref{REM constant f} ). We omit
the details for lack of space.} 
\end{proof}

For what concerns $\strictFACQplus{}$ queries, the following result can be proved.

\begin{theorem}~\label{RES strictFACQplus}
The query problem $\strictFACQplus{}$ can be solved in time
$O(|\varphi|.|D|)$
\end{theorem}

\begin{proof}
The proof, that is a generalization of the proof for
$\strictFACQ{}$, is similar and, in several aspects, is simpler
than that of the similar result for $\strictFACQineq{}$. Let us
mention essentially the differences. W.l.o.g., let $\varphi\in
\strictFACQplus{}$ be a formula of the form

\[
\varphi (x) \equiv \exists y_1 \dots \exists y_{d-1} \exists y_d :
\Psi(y_0,y_1,\dots,y_{d-1}) \wedge \tu u (y_{i_0}) = \tu v (y_d)
\wedge f(y_{i_0}) \theta g(y_d) \wedge \gamma(y_d)
\]

\noindent where $y_0$ is $x$, $\theta \in \{\neq, <, \leq, >,
\geq\}$, $0\leq i_0 \leq d-1$, $\gamma(y_d)$ is a quantifier-free
formula on the unique variable $y_d$, and $\tu u (y_{i_0}) = \tu v
(y_d)$ stands for $\bigwedge_{j\leq l} u_j (y_{i_0}) = v_j (y_d)$.
Formula $\varphi$ can be equivalently written as:

\[
\varphi (x) \equiv \exists y_1 \dots \exists y_{d-1} : \Psi(\tu y)
\wedge \delta (y_{i_0})
\]

\noindent where $\tu y = (y_0,y_1,\dots,y_{d-1})$ and $\delta$ is
the following two-variable formula:

\[
\delta(y) \equiv \exists z : \gamma(z) \wedge \tu u (y) = \tu v
(z) \wedge f(y) \theta g(z)
\]

The key point is the following:

\begin{lemma}~\label{RES lemma for strictFACQplus}
The set $D_0 = \delta(\calF) = \{a \in D: (\calF, a) \models
\delta (y)\}$ is computable in time $O(|\delta|.|D|)$
\end{lemma}

\begin{proof}
The set $B = \gamma(\calF) = \{b \in D: (\calF, b) \models \gamma
(z)\}$ is obviously computable in time $O(|\gamma|.|D|)$. Assume
that the comparison symbol $\theta$ is $<$ (the other cases are
variants of this case). Now, compute and lexicographically sort
the following lists of $(l+3)$-tuples (in time $O(l.|D|)$):

\[
Y = \{(\tu u (y), f(y), 1, y): y \in D\}, \mbox{ and }
\]

\[
Z = \{(\tu v (z), g(z), 0, z): z \in B\}.
\]

Then, merge the sorted lists $Y$, $Z$ into the sorted list $L$. It
is easy to see that the following fact holds:

\begin{fact}~\label{FACT L}
$\delta(\calF)$ is the set of elements $y\in D$ such that there
exists $z\in B$ such that $\tu u(y) = \tu v(z) $ and $(\tu u (y),
f(y), 1, y)$ occurs before $(\tu v (z), g(z), 0, z)$ in $L$.
\end{fact}

Using this fact, the following algorithm computes $\delta(\calF)$
(knowing set $B$) in time $O(l.|D|)$.

\begin{itemize}

\item Partition the sorted list $L$ into nonempty (sorted) sublists
$L(\tu a)$, for $\tu a \in \tu u(D) \cup \tu v(B)$, according to
the first $l$-tuple $\tu u(y) = \tu a$ or $\tu v(z) = \tu a$ of
each tuple.

\item In each sorted list $L(\tu a)$, compute the last tuple,
denoted by $Max_B(\tu a)$, of the form $(\tu v(z) = \tu a, g(z),
0, z)$, $z\in B$, with $g(z)$ maximal if such element exists.
Otherwise, set $Max_B(\tu a) = - \infty$.

\item In each $L(\tu a)$, compute the list $L^<(\tu a)$ of the
tuples of the form $(\tu u(y) = \tu a, f(y), 1, y)$ that occur
before $Max_B(\tu a)$ in $L(\tu a)$. By convention, $L^<(\tu a)$
is empty in case $Max_B(\tu a) = - \infty$.

\item Return the set of elements $y$ that appear in the lists $L^<(\tu a)$. By Fact~\ref{FACT
L}, this is clearly the required set $\delta(\calF)$.

\end{itemize}

Globally, $\delta(\calF)$ is computed in time $O((|\gamma| +
l).|D|) = O(|\delta|.|D|)$. This proves the lemma.
\end{proof}

\noindent \textit{End of proof of Theorem~\ref{RES
strictFACQplus}: \ } Let $\calF'$ be the expansion of structure
$\calF$ defined as $\calF' = (\calF, D_0)$ where $D_0$ is the
unary predicate defined as $D_0 = \delta(\calF)$. Let $\varphi'$
denote the following formula, of signature expanded with $D_0$:

\[
\varphi'(x) \equiv \exists y_1 \dots \exists y_{d-1} :
\Psi(x,y_1,\dots,y_{d-1}) \wedge D(y_{i_0})
\]

\noindent where $y_{i_0}\in \{x,y_1,\dots, y_{d-1}\}$. By
construction, we have:

\begin{fact}~\label{FACT strictFACQplus 2}
$\varphi (\calF) = \varphi'(\calF')$.
\end{fact}

In order to simply compare the lengths of $\varphi$ and
$\varphi'$, let us introduce a simplified notion of formula
length: let $|\varphi|_s$ denote the number of quantifiers of
$\varphi$ plus its number of occurrences of atoms. Clearly, it
holds: $|\varphi| = \Theta(|\varphi|_s)$. By construction, we get
the following fact:

\begin{fact}~\label{FACT strictFACQplus 3}
$|\varphi'|_s = |\varphi|_s - |\delta|_s + 1$.
\end{fact}

Lemma~\ref{RES lemma for strictFACQplus} immediately yields the
following:

\begin{fact}~\label{FACT strictFACQplus 4}
The expansion $\calF \mapsto \calF'$, i.e., the computation of the
added unary relation $D_0= \delta(\calF)$ is computed in time
$0(|\delta|_s.|D|)$.
\end{fact}

Iterating the transformation $(\calF, \varphi) \mapsto
(\calF',\varphi')$ $d$ times allows to eliminate successively the
quantifed variables $y_d, y_{d-1}, \dots, y_1$; this can be
performed in total time $O((|\varphi|_s + d).|D|)$ by
Facts~\ref{FACT strictFACQplus 3} and~\ref{FACT strictFACQplus 4},
and hence in time $O(|\varphi|.|D|)$ as required. This completes
the proof of the theorem.
\end{proof}

\begin{remark}~\label{REM more than one comparison}
Allowing more than one comparison along the edges of the tree decomposition leads to a class of queries that seems intrinsically non-linear. Let's consider the very simple following formula with two comparisons:

\[
\exists x \exists y : \ f_1(x) \leq g_1(y) \wedge f_2(x) \leq g_2(y) .
\]

Finding two satisfying witnesses $x$ and $y$, amounts to find   lexicographically ordered pairs $(f_1(x),f_2(x))$ and $(g_1(y),g_2(y))$ which seems not doable in linear time (even if "tables" $(f_1,f_2)$ and $(g_1,g_2)$ are already sorted). 
\end{remark}

The following theorem states the complexity of our (functional)
acyclic queries in the general case.

\begin{theorem}~\label{RES functional generalized query}
The $\FACQineq{}$ (resp. $\FACQ{}$, $\FACQplus{}$) query problem
can be solved in time $f(|\varphi|).|D|.|\varphi(\calF)|$ (resp.
$O(|\varphi|.|D|.|\varphi(\calF)|)$) for some function $f$.
\end{theorem}

\begin{proof} We prove that,
for any function $f:\mathbb{N} \mapsto \mathbb{R^+}$, if problem
$\strictFACQplus{}$ (resp. $\strictFACQineq{}$) can be solved in
time $f(|\varphi|).|D|$ then problem $\FACQplus{}$ (resp.
$\FACQineq{}$) can be solved in time
$f(|\varphi|).|D|.|\varphi(\calF)|$ for the same function $f$.
Combined with Theorem~\ref{RES strictFACQineq and strictFACQ}
and~\ref{RES strictFACQplus}, this yields the desired result.

Let $\calF$ be a functional structure and $\varphi(x_1,\dots,x_k)$
be a formula for the query problem $\FACQplus{}$ or $\FACQineq{}$. For
$i=1,\dots, k$, let:

\[
E_i = \{(x_1,\dots,x_i)\in D^i: \ (\calF,x_1,\dots,x_i) \models
\exists x_{i+1} \dots \exists x_k \varphi\}
\]

\noindent Obviously, $\varphi(\calF)=E_k$. Sets $E_1$, \dots, $E_k$ are
computed inductively by the following algorithm that only evaluates
strict acyclic queries as subroutines.

\bigskip


\medskip

 \noindent $E_1 \leftarrow \{x_1 \in D: (\calF,x_1) \models
\exists x_{2} \dots \exists x_k \varphi\}$

\noindent \textbf{For $i$ from $2$ to $k$ do}

$E_i \leftarrow \emptyset$

\textbf{For all $(x_1,\dots,x_{i-1})\in E_{i-1}$ do}

\ \ $S \leftarrow \{ x_i\in D: \ (\calF,x_1,\dots,x_i) \models
\exists x_{i+1} \dots \exists x_k \varphi\}$ \hskip 1cm (*)

\ \ $E_i \leftarrow E_i \cup \{(x_1,\dots,x_{i-1},x_i): \ x_i \in
S \}$

\textbf{End}

\noindent \textbf{End}

\noindent $\varphi(\calF) \leftarrow E_k$

\bigskip

\noindent The main step of the algorithm, that is step (*), requires
 time $f(|\varphi|).|D|$. It is repeated, a number of times
bounded by:

\[
card(E_1)+card(E_2)+\dots + card(E_k) \leq
k.card(E_k)=|E_k|=|\varphi(\calF)|
\]

\noindent This yields total time $f(|\varphi|).|D|.|\varphi(\calF)|$.
\end{proof}

\bigskip

Finally, let us give another consequence of our results in the
functional setting. Any two-variable quantifier-free (CNF)
functional formula $\psi(x,y)$ is acyclic because any undirected
graph with at most two vertices is acyclic. Let $\FFOvar2{}$
denote the set of functional first-order formulas (not necessarily
in prenex form) with only two variables $x,y$ which may be
quantified several times. Denote by $\strictFFOvar2{}$ its
restriction to strict queries.

\begin{corollary}\label{RES two var}
The $\strictFFOvar2{}$  query problem is computable in time
$O_{\varphi}(|\calF|)$.
\end{corollary}

\begin{proof}
The proof is done by induction on the structure (i.e.,
subformulas) of the input formula $\varphi$ by using
Theorem~\ref{RES: FO acyclic}.
\end{proof}

\section{Application to the complexity of relational acyclic queries}

In the context of "classical", i.e., relational conjunctive queries,
Theorem~\ref{RES functional generalized query}  immediately yields
the following improvement of the time bound
$g(|Q|).|\textbf{db}|.|Q(\textbf{db})|. \log^2 |\textbf{db}|$
(for some function $g$)  proved by~\cite{PapadimitriouY-99} for
the complexity of acyclic queries with inequalities.

\begin{corollary}~\label{RES generalized query} The $\ACQineq{}$
 (resp. $\strictACQineq{}$) query problem can be solved in time
 $f(|Q|).|\textbf{db}|.|Q(\textbf{db})|$ (resp. $f(|Q|).|\textbf{db}|$)
where $Q$ is the input query and \textbf{db} is the input database.
\end{corollary}

\begin{proof}
This comes from Theorem~\ref{RES functional generalized query} and
from the fact that the class $\ACQineq{}$ can be linearly
interpreted by the class  $\FACQineq{}$ (see section~\ref{SEC From
CQ to FQ}).
\end{proof}

\bigskip

Another consequence of Theorems~\ref{RES strictFACQineq and
strictFACQ} and~\ref{RES functional generalized query} is  an
alternative proof of the following well-known result of
\cite{Yannakakis-81} (see also~\cite{FlumFG-02}) that we slightly
generalize since now also restricted comparisons are allowed.

\begin{corollary}
The $\ACQ{}$ and  $\ACQplus{}$ (resp. $\strictACQ{}$ and $\strictACQplus{}$) query problems can be solved in time $O(|Q|.|\textbf{db}|.|Q(\textbf{db})|)$ (resp.
$O(|Q|.|\textbf{db}|)$).
\end{corollary}

In a two-atom query  each database predicate appears at most two times. These kind of queries have been studied in~\cite{KolaitisV-00,Saraiya-91} mainly in the context of query-containment.  
A consequence of Corollary~\ref{RES two var}, is the following.

\begin{corollary}\label{RES two atom}
Any two-atom conjunctive query with inequalities can be evaluated in time 
$O_{\varphi}(|\textbf{db}|)$ i.e. in time $O_{\varphi}(|T_1|+|T_2|)$ where $T_1$ and $T_2$ are the two input tables.
\end{corollary}

\section{Enumeration of query results}~\label{SEC enumeration}

For all kind of queries considered in this paper, the complexity of the evaluation process can be done in time $f(|Q|).|\textbf{db}|.|Q(\textbf{db})|$. In other words, coming back to data complexity, this is equivalent to say that there exists a polynomial total time algorithm (in the size of the input and the output) that generates the output tuples. It is natural to ask whether one can say more on the efficiency of this enumeration process. This could be justified, for example, in situation where only parts of the results are really needed quickly or when having solutions one by one but regularly is required (e.g., in order to be tested by an other procedure that runs in parallel). Some remarks on this subject are sketched in this section.

One of the most widely accepted notion of tractability in the context of generation of solutions  is the following. A problem $P$ is said to be solvable within a \textit{Polynomial (resp. linear) Delay} if there exists an algorithm that outputs a first solution in polynomial (resp. linear) time  (in the size of the input only) and generates all solutions of $P$ with a polynomial (resp. linear) delay between two consecutives ones (see~\cite{JohnsonYP-88} for an introduction to complexity measures for enumeration problems). Of course, a \textit{Polynomial Delay} algorithm is polynomial total time (but, unless surprise, the converse is not true).

Not too surprisingly, our complexity results can be adapted to obtain polynomial, even linear, delay algorithms for acyclic queries as shown by the following corollary. 

\begin{corollary}~\label{COR enumeration}
Generating all results of a $\FACQineq{}$ (resp. $\FACQ{}$, $\FACQplus{}$, $\ACQineq{}$, $\ACQ{}$, $\ACQplus{}$) query can be done with a linear delay (and with linear space also).
\end{corollary}

\begin{proof}
We proceed in a similar way as for Theorem~\ref{RES functional generalized query}. Results for relational query classes are obtained by reduction. Let $\calF$ be a functional structure and $\varphi(x_1,\dots,x_k)$ be a functional query in $\FACQineq{}$, $\FACQ{}$ or $\FACQplus{}$. 

The simple (recursive) algorithm below outputs all satisfying tuples of $\varphi(x_1,\dots,x_k)$.

 \begin{algorithm}
\caption{ Eval($i,\varphi(x_i,\dots,x_k),\calF,sol$)}
\begin{algorithmic}[0]
\If{$i=k+1$} 
	\State \textbf{Output} $sol$ 
\EndIf
\State $E_i \leftarrow \{ x_i\in D: (\calF,x_i) \models
        \exists x_{i+1} \dots \exists x_k \varphi\}$~\label{Cptr}
\For{$a \in E_i$}
	\State $sol \leftarrow (sol,a)$
	\State $\varphi \leftarrow \varphi(x_i/a,x_{i+1},\dots,x_k)$
	\State  Eval($i+1,\varphi,\calF,sol$)
 \EndFor
\end{algorithmic}
\end{algorithm}

Due to results of the preceding sections, computing $E_i$ can be done, in all cases, in time $f(|\varphi|.|D|)$. Then, running \textit{Eval($1,\varphi(x_1,\dots,x_k),\calF,\emptyset$)}  generates all solutions $sol$ in a depth-first manner with a linear delay detween each of them. It can be easily rewritten in a sequential way to use linear space.   
\end{proof}

\section{Fixed-parameter linearity of some natural problems}~\label{SEC Fixed-parameter linearity of some natural problems}

In this part of the paper, the different kind of formulas introduced so far are used to define classical algorithm properties as query problems. This method provides a simple and uniform method to cope with the complexity of these problems. In all cases, the complexity bound found with this method reaches or improve  the best bound known so far (at least in terms of data complexity). However, some of these problems have been the object of intensive researches and recent optimize ad-hoc algorithms (against which a general and uniform method can not compete) have better constant values.

\subsection{Acyclic Subgraph problems}

Given two graphs $G=\st{V}{E}$ and $H=\st{V_H}{E_H}$, $H$ is said to be a \textit{subgraph} (resp. \textit{induced subgraph}) of $G$ if there is a one-to-one function $g$ from $V_H$ to $V$ such that, for all $u,v\in V_H$, $E(g(u),g(v))$ if (resp. if and only if) $E(u,v)$. Also, a graph $G$ is of maximum degree $d$ if none of its vertex belongs to more than $d$ edges. This gives rise to the two following problems. 

\bigskip

\noindent \pb{acyclic subgraph isomorphism (a.s.i.)}\\
\noindent \begin{tabular}{rl}
\textbf{Input:} & \parbox[t]{300 pt}{an acyclic graph $H$ and a graph $G$}\\
\textbf{Parameter:} & $|H|$.\\
\textbf{Question:} & is $H$ a subgraph of $G$ ?\\
\end{tabular}

\bigskip

\noindent \pb{acyclic induced subgraph isomorphism (a.i.s.i.)}\\
\noindent \begin{tabular}{rl}
\textbf{Input:} & \parbox[t]{300 pt}{an acyclic graph $H$ and a graph $G$ of maximum degree $d$}\\
\textbf{Parameter:} & $|H|, d$.\\
\textbf{Question:} & is $H$ an induced subgraph of $G$ ?\\
\end{tabular}

\bigskip

The treewidth of a graph $G$ is the maximal size of a node in a tree decomposition of $G$.    
In~\cite{PlehnV-90} it is proved that for graphs $H$ of treewidth at most $w$, testing is $H$ is a subgraph (resp. induced subgraph) of $G$ can be done in time $f(|H|).|G|^{w+1}$ (resp. $f(|H|,d).|G|^{w+1}$). For the particular case of acyclic graphs (which have tree width $1$), the bounds given in~\cite{PlehnV-90} can be improved.  The following corollary is easily obtained from our results.

\begin{corollary}\label{COR ASI} The two following results hold:

\begin{itemize}
\item Problem \pb{a.s.i.} can be solved in time $f(|H|).|G|$.

\item Problem \pb{a.i.s.i.} can be solved in time $f(|H|,d).|G|$.
\end{itemize}

For the two problems, generating all satisfying subgraphs can be done with a linear delay. 
\end{corollary}

\begin{proof}
We will express problem $\pb{a.s.i.}$ as a boolean $\ACQineq{}$ query.
Let $G=\st{V}{E}$, $H=\st{V_H=\{h_1,\dots,h_k\}}{E_H}$ be the two input graphs.
Let $Q$ be the following formula:

\[
Q \equiv
\exists x_1 \dots \exists x_k : \ \bigwedge_{i,j\leq k} x_i \neq x_j \wedge \bigwedge_{E_H(h_i,h_j)} E(x_i,x_j)
\] 

Since $H$ is acyclic, formula $Q$ defines an $\ACQineq{}$ query whose size is linear in the size of the graph $H$.
It is easily seen that $Q$ is true in $G$ if and only if it admits $H$ as a subgraph. The complexity bound follows from Corollary~\ref{RES generalized query}. 

\bigskip

For problem $\pb{a.i.s.i.}$, let again $G$ and $H=\st{V_H=\{x_1,\dots,x_k\}}{E_H}$ be the two inputs of the problem.
Since $^G$ is of maximum degree $d$, we partition its vertex set $V$ into $d$ sets $V^1,\dots,V^d$ where each $V^{\alpha}$ contains vertex of degree $\alpha$.  This can be done in linear time from $G$. We proceed the same for graph $H$ and obtain the sets $V_H^1, \dots, V_H^d$. In case there exists a vertex in $H$ of degree greater than $d$, it can be concluded immediately that the problem has no solution. Now, let $Q$ be the following formula:

\[
Q \equiv
\exists x_1 \dots \exists x_k : \ \bigwedge_{i,j\leq k} x_i \neq x_j \wedge \bigwedge_{V_H^{\alpha}(h_i)} V_H^{\alpha}(x_i) \wedge \bigwedge_{E_H(h_i,h_j)} E(x_i,x_j).
\] 

Formula $Q$ simply check that $H$ is a subgraph of $G$ and that each distinguished vertex $x_i$ has the same degree than its associated vertex $h_i$ of $H$. The size of $Q$ is linear in the size of $H$ and $d$.
Again, $Q$ defines a boolean $\ACQineq{}$ query and the result follows again from Corollary~\ref{RES generalized query}.
The bound on the linear delay comes from Corollary~\ref{COR enumeration}.
\end{proof}

\subsection{Covering and matching problems}

\noindent \pb{multidimensional matching}\\
\noindent \begin{tabular}{rl}
\textbf{Input:} & \parbox[t]{300 pt}{a set $M \subseteq X_1 \times \dots \times X_r$ where the $X_i$ are pairwise disjoints}\\
\textbf{Parameter:} & $r, k$.\\
\textbf{Question:} & \parbox[t]{300 pt}{is there a subset $M' \subseteq M$ with $|M'|=k$, such that no two elements of $M'$ agree in any coordinate ?}\\
\end{tabular}

\bigskip

\begin{corollary}\label{COR MDM} 
Problem \pb{multidimensional matching} can be solved in time $O_{r,k}(|M|)$.
\end{corollary}

\begin{proof}
Let $\calF_M = \st{M}{f_1,\dots,f_r}$ where for all $x=(x_1,\dots, x_r)\in M$, it is set $f_i(x)=x_i$. Then, there exists a multidimensional matching $M'$ of $M$ if and only if:

\[
\calF_M \models \exists x_1 \dots \exists x_k : \ \bigwedge_{i\leq r} \bigwedge_{1\leq j < h \leq k} f_i(x_j) \neq f_i(x_h)
\] 
\end{proof}

Corollary~\ref{COR MDM} below  improves the bound of $O_{r,k}(|M|(\log |M|)^6)$  (reported in~\cite{DowneyF-99}) obtained by perfect hashing methods. A recent result however of~\cite{FellowsETal-04} based on the color coding method of~\cite{AlonYZ-95}  gives a bound of $O(|M| + 2^{O(k)})$ for the $r$-\pb{multidimensional matching} problem.

\medskip

The following problems are also known to  
be fixed-parameter tractable~\cite{DowneyF-99}.

\bigskip

\noindent \pb{unique hitting set}\\
\noindent \begin{tabular}{rl}
\textbf{Input:} & \parbox[t]{300 pt}{a set $X$ and $k$ subsets $X_1,\dots, X_k$ of $X$.}\\
\textbf{Parameter:} & $k$.\\
\textbf{Question:} & \parbox[t]{305 pt}{is there a set $S\subseteq X$ such that for all $i$, $1\leq i \leq k$, $|S \cap X_i| = 1$ ?}\\
\end{tabular}

\medskip

\noindent \pb{antichain of $r$-subsets}\\
\noindent \begin{tabular}{rl}
\textbf{Input:} & \parbox[t]{300 pt}{a collection $\calF$ of $r$ subsets of a set $X$, a positive integer $k$.}\\
\textbf{Parameter:} & $r,k$.\\
\textbf{Question:} & \parbox[t]{300 pt}{are there $k$ subsets $S_1,\dots,S_k\in \calF$ such that $\forall i,j\in \{1,\dots,k\}$ with $i\neq j$, both $S_i - S_j$ and $S_j - S_i$ are nonempty  ?}\\
\end{tabular}

\medskip

\noindent \pb{disjoint $r$-subsets}\\
\noindent \begin{tabular}{rl}
\textbf{Input:} & \parbox[t]{300 pt}{a collection $\calF$ of $r$ subsets of a set $X$, a positive integer $k$.}\\
\textbf{Parameter:} & $r,k$.\\
\textbf{Question:} & \parbox[t]{300 pt}{are there $k$ disjoint subsets of $\calF$ ?}\\
\end{tabular}

\medskip

\begin{corollary}\label{COR MDM} 
Problems \pb{unique hitting set}, \pb{antichain of $r$-subsets} and \pb{disjoint $r$-subsets} can be solved in time $O_{r,k}(|M|)$. In all cases, the respective sets of solutions can be generated with a linear delay.
\end{corollary}

\begin{proof}
The following acyclic formula holds for the \pb{unique hitting set} problem:

\[
\varphi \equiv \exists x_1 \dots \exists x_k : \ \bigwedge_{i\leq k} X_i(x_i) \bigwedge_{1\leq i < j \leq k} (x_i \neq x_j \Rightarrow \neg X_j(x_i)). 
\] 
 
The formulas are similar for the two other problems. 
\end{proof}

\section{Conclusion: summary of results and open problems}

The following array summarizes the main results of this
paper.
For all our classes of ("classical", i.e., relational, or functional) queries we make use
here of the notation $\varphi$ for the query formula, $\calS$ for
the database, i.e., the input structure $\textbf{db}$ or $\calF$, and $\varphi(\calS)$ for the
result of the query, i.e., the output.

\[
\begin{tabular}{|c|c|}\hline
  \textbf{Query Problems} & \textbf{Complexity} \\
  \hline \hline
   $\strictACQ{},  \strictACQplus{}, \strictFACQ{}, \strictFACQplus{}$ & $|\varphi|.|\calS|$
   \\ \hline
   $\strictACQineq{}, \strictFACQineq{}, \strictFFOvar2{}$ & $f(|\varphi|).|\calS|$ \\ \hline
 $\ACQ{}, \ACQplus{}, \FACQ{},  \FACQplus{}$  &   $|\varphi|.|\calS|.|\varphi(\calS)|$ \\ \hline
 $\ACQineq{}, \FACQineq{}$ &
$f(|\varphi|).|\calS|.|\varphi(\calS)|$ \\ \hline
\end{tabular}
\]

\noindent Note that among those complexity results the only ones
to be known before this paper (to our knowledge) where those
concerning  $\ACQ{}$ and $\strictACQ{}$.

We are convinced that (variants
of) our technics of construction of \textit{minimal samples} can
be efficiently implemented to compute such queries. The reason is
that we think that  the total number of minimal samples should be
very low in most databases.

Finally, four lines of research are worthwhile to develop:

\begin{itemize}

\item Generalize our complexity results to tractable or f.p.
tractable tree-like queries e.g., queries of bounded tree-width
(see~\cite{ChekuriR-00, FlumFG-02}) or of bounded hypertree-width
(\cite{GottlobLS-02}). Our reduction technic from relational to functional queries, which preserves acyclicity, may permit also to control the value of the tree-width when passing from one context to the other.

\item Apply our results to constraint satisfaction problems by
using the now well-known correspondence between conjunctive query
problems and constraint problems (see among
others~\cite{KolaitisV-00}).


\item Enlarge the classes of tractable or f.p. tractable problems as
much as possible, i.e., determine the frontier of
tractability/intractability, and obtain for the (f.p.) tractable
problems the best sequential or parallel algorithms; e.g., it is
reasonable to conjecture that the $\ACQplus{}$ evaluation problem
is highly parallelizable as it is known for $\ACQ{}$ (see for
example~\cite{GottlobLS-01}
 which proves that this problem is LOGCFL-complete).

\item Apply our methods to queries over tree-structured data
(recall that a rooted tree can be seen as a graph of a unary
function).

 \end{itemize}


\bibliographystyle{alpha}
\bibliography{../../../biblio/perso,../../../biblio/central,../../../biblio/mabibli}

\newcommand{\etalchar}[1]{$^{#1}$}
\begin{thebibliography}{FKN{\etalchar{+}}04}

\bibitem[AHU74]{AhoHU-74}
A.~V. Aho, J.~E. Hopcroft, and J.~D. Ullman.
\newblock {\em The Design and Analysis of Computer Algorithms}.
\newblock Addison-Wesley, 1974.

\bibitem[AHV95]{AbiteboulHV-95}
S.~Abiteboul, R.~Hull, and V.~Vianu.
\newblock {\em Foundation of databases}.
\newblock Addison-Wesley, 1995.

\bibitem[AYZ95]{AlonYZ-95}
N.~Alon, R.~Yuster, and U.~Zwick.
\newblock Color coding.
\newblock {\em Journal of the ACM}, 42(4):844--856, 1995.

\bibitem[CM77]{ChandraM-77}
A.K. Chandra and P.M. Merlin.
\newblock Optimal implementation of conjunctive queries in relational
  databases.
\newblock In ACM~New York, editor, {\em Proceedings of the 9th Annual ACM
  Symposium on Theory of Computing}, pages 77--90, 1977.

\bibitem[CR00]{ChekuriR-00}
C.~Chekuri and A.~Rajaraman.
\newblock Conjunctive query containment revisited.
\newblock {\em Theoretical Computer Science}, 239(2):211--229, 2000.

\bibitem[DF99]{DowneyF-99}
R.~G. Downey and M.~R. Fellows.
\newblock {\em Parametrized complexity}.
\newblock Springer-Verlag, 1999.

\bibitem[DGO04]{DurandGO-04}
A.~Durand, E.~Grandjean, and F.~Olive.
\newblock New results on arity vs. number of variables.
\newblock Research report 20-2004, LIF, Marseille, France, April 2004.

\bibitem[EF99]{EbbinghausF-99}
H.-D. Ebbinghaus and J.~Flum.
\newblock {\em Finite Model Theory}.
\newblock Springer-Verlag, 2nd edition, 1999.

\bibitem[Fag83]{Fagin-83}
R.~Fagin.
\newblock Degrees of acyclicity for hypergraphs and relational database
  schemes.
\newblock {\em Journal of the ACM}, 30(3):514--550, 1983.

\bibitem[FFG02]{FlumFG-02}
J.~Flum, M.~Frick, and M.~Grohe.
\newblock Query evaluation via tree decompositions.
\newblock {\em Journal of the ACM}, 49(6):716--752, 2002.

\bibitem[FKN{\etalchar{+}}04]{FellowsETal-04}
M.~Fellows, C.~Knauer, N.~Nishimura, P.~Ragde, F.~Rosamond, U.~Stege,
  D.~Thilikos, and S.~Whitesides.
\newblock Faster fixed-parameter tractable algorithms for matching and packing
  problems.
\newblock In {\em European Symposium on Algorithms 2004}, pages 311--322, 2004.

\bibitem[GLS01]{GottlobLS-01}
G.~Gottlob, N.~Leone, and F.~Scarcello.
\newblock The complexity of acyclic conjunctive queries.
\newblock {\em Journal of the ACM}, 48(3):431--498, 2001.

\bibitem[GLS02]{GottlobLS-02}
G.~Gottlob, N.~Leone, and F.~Scarcello.
\newblock Hypertree decompositions and tractable queries.
\newblock {\em Journal of Computer and System Sciences}, 64(3):579--627, 2002.

\bibitem[GO04]{GrandjeanO-04}
E.~Grandjean and F.~Olive.
\newblock Graphs properties checkable in linear time in the number of vertices.
\newblock {\em Journal of Computer and System Sciences}, 68(3):546--597, 2004.

\bibitem[Gra79]{Graham-79}
R.~Graham.
\newblock On the universal relation.
\newblock Technical report, Univ. Toronto, 1979.

\bibitem[GS02]{GrandjeanS-02}
E.~Grandjean and T.~Schwentick.
\newblock Machine-independent characterizations and complete problems for
  deterministic linear time.
\newblock {\em SIAM Journal on Computing}, 32(1):196--230, 2002.

\bibitem[JYP88]{JohnsonYP-88}
D.~S. Johnson, M.~Yannakakis, and C.~H. Papadimitriou.
\newblock On generating all maximal independent sets.
\newblock {\em Information Processing Letters}, 27(3):119--123, 1988.

\bibitem[KV00]{KolaitisV-00}
P.~G. Kolaitis and M.~Y. Vardi.
\newblock Conjunctive-query containment and constraint satisfaction.
\newblock {\em Journal of Computer and System Science}, 61(2):302--332, 2000.

\bibitem[Lib04]{Libkin-04}
L.~Libkin.
\newblock {\em Elements of finite model theory}.
\newblock EATCS Series. Springer, 2004.

\bibitem[PV90]{PlehnV-90}
J.~Plehn and B.~Voigt.
\newblock Finding minimally weighted subgraphs.
\newblock In Springer, editor, {\em 16th workshop on graph theoretic concepts
  in computer science}, volume 484 of {\em Lecture Notes in Computer Science},
  pages 18--29, 1990.

\bibitem[PY99]{PapadimitriouY-99}
C.~Papadimitriou and M.~Yannakakis.
\newblock On the complexity of database queries.
\newblock {\em Journal of Computer and System Sciences}, 58(3):407--427, 1999.

\bibitem[Sar91]{Saraiya-91}
Y.~Saraiya.
\newblock {\em Subtree elimination algorithms in deductive databases}.
\newblock PhD thesis, Stanford University, 1991.

\bibitem[Ull89]{Ullman-89}
J.D. Ullman.
\newblock {\em Principles of Database and Knowledge-Base Systems, Volume II}.
\newblock Computer Science Press, 1989.

\bibitem[Yan81]{Yannakakis-81}
M.~Yannakakis.
\newblock Algorithms for acyclic database schemes.
\newblock In {\em Proceedings of the 7th International Conference on Very Large
  Databases}, pages 82--94, 1981.

\bibitem[YO79]{YuO-79}
C.~T. Yu and M.~Z \"{O}zsoyoglu.
\newblock An algorithm for tree-query membership of a distributed query.
\newblock In IEEE Computer~Society Press, editor, {\em IEEE COMPSAC}, pages
  306--312, 1979.

\end{thebibliography}
\nocite{DowneyF-99, AbiteboulHV-95, GottlobLS-01, Libkin-04,
Ullman-89}

\end{document}